\documentclass[4pt,twocolumn]{article}
\pdfoutput=1
\usepackage[lmargin=0.75in,rmargin=0.75in,tmargin=1.0in,bmargin=1.0in]{geometry}
\usepackage{wrapfig}

\usepackage{graphicx}
\usepackage{pdfpages}

\usepackage[font={small}]{caption} 

\usepackage[square,numbers,sort]{natbib}
\usepackage{authblk}

\begin{document}
\title{Laboratory measurements of HDO/H$_{2}$O isotopic fractionation during ice deposition in simulated cirrus clouds}
\date{}
\setcounter{Maxaffil}{2}
\author[1]{Kara Lamb} 
\author[1]{Ben Clouser}
\author[2]{Maximilien Bolot} 
\author[2]{Laszlo Sarkozy}
\author[3]{Volker Ebert}
\author[4]{Harald Saathoff} 
\author[4]{Ottmar M\"ohler} 
\author[2]{Elisabeth Moyer}

\affil[1]{Dept.\ of Physics, University of Chicago}
\affil[2]{Dept.\ of the Geophysical Sciences, University of Chicago}
\affil[3]{Physikalisch-Technische Bundesanstalt }
\affil[4]{Institute for Meteorology and Climate Research, Karlsruhe Institute of Technology}

\maketitle

\begin{abstract}
\textbf{The stable isotopologues of water have been used in atmospheric and climatic studies for over fifty years, because the temperature-dependent preferential condensation of heavy isotopologues during phase changes makes them useful diagnostics of the hydrological cycle. However, the degree of preferential condensation has never been directly measured at temperatures below 233 K ($-40^{\circ}$C), conditions necessary for cirrus formation in the atmosphere and routinely observed at surface elevation in polar regions. (Models generally assume an extrapolation from the warmer experiments of Merlivat and Nief, 1967 \cite{Merlivat67}.) 
Non-equilibrium effects that should alter preferential partitioning have also not been well-characterized experimentally \cite{Jouzel84}.
We present here the first direct experimental measurements of the HDO/H$_2$O equilibrium fractionation factor between vapour and ice ($\alpha_{\mathrm {eq}}$) at cirrus-relevant temperatures, and the first quantitative validation of the kinetic modification to equilibrium fractionation expected to occur in supersaturated conditions.
In measurements of the evolving isotopic composition of water vapour during cirrus formation experiments in the AIDA chamber, we find $\alpha_{\mathrm{eq}}$ several percent lower than has been assumed. 
In a subset of diffusion-limited experiments, we show that kinetic isotope effects are consistent with published models \cite{Jouzel84,Nelson11}, including allowing for small surface kinetic effects.
These results are significant for the inference of cirrus and convective processes from water isotopic measurements.}
\end{abstract}

Accurate values of the vapour-ice fractionation factor are needed in many isotopically-based paleoclimate or atmospheric studies: for paleotemperature or paleoaltimetry reconstructions with process-based models \cite{Jouzel96}, 
for characterising the hydrological cycle \cite{Bony08,Gedzelman94,Risi08}, and for 
diagnosing convective transport of water to the tropical tropopause layer (TTL) \cite{Moyer96,Hanisco07,Randel12,Blossey10}. 
To date, water isotopologues have been introduced into 12 general circulation models for use as atmospheric tracers \citep[e.g.][]{Joussaume84,Jouzel87a,Lee08}. For the HDO/H$_2$O system, all use extrapolations of $\alpha_{\mathrm{eq}}$ from Merlivat and Nief 1967 (henceforth M67), made at temperatures warmer than the regime for cirrus formation \cite{Merlivat67}. 

Measuring $\alpha_{\mathrm{eq}}$  at cold temperatures is difficult largely because water vapour pressure becomes so small: in the cold uppermost troposphere, mixing ratios of H$_2$O can be a few ppm and HDO a few ppb.  Equilibrium fractionation in fact becomes very large in these conditions,  because the effect scales approximately as the mass ratio of substituted atoms (D:H = 2) \cite{Urey47} and rises as $\sim$ 1/T$^2$: the temperature dependence is typically assumed as 
\begin{equation}
 \alpha_{\mathrm{eq}}\left(T\right)=\exp{\left(a_{0}+\frac{a_{1}}{T^{2}}\right)}, \label{eq:tempdepend}
 \end{equation}
the high-temperature limit of \cite{Criss91, Urey47}. In M67, $\alpha_{\mathrm{eq}}$ exceeds 1.4 ($>$40\% HDO enhancement in ice) at 190 K, the largest vapour pressure isotope effect seen in natural systems. 
In 2013, Elleh\o j et al.\ (henceforth E13) reported measurements suggesting still stronger fractionation, 15\% higher at 190 K \cite{Ellehoj13}. (See supplementary materials S1 for all previous estimates.) Differences on this order would alter interpretation of isotopic profiles, e.g.\ in terms of the balance of dehydration vs.\ convective sources \cite{Bolot13}. 

Potential kinetic modifications to this fractionation are poorly characterized by experimental studies. Jouzel and Merlivat 1984 \cite{Jouzel84} explained non-equilibrium isotopic signatures in polar snow as the result of reduced effective fractionation when ice grows in diffusion-limited (supersaturated) conditions: preferential uptake should isotopically lighten the near-field vapour around growing ice crystals, with the effect enhanced by the fact that heavier isotopologues have lower diffusivity. The kinetic modification factor $\alpha_{k}$ can then be written in terms of properties of the bulk gas:
\begin{equation}
\alpha_{k}=\frac{S_{i}}{\alpha_{\mathrm{eq}}\cdot\mathrm{d}\left(S_{i}-1\right)+1}
\label{eq:alphak}
\end{equation}
where $\mathrm{d}$ is the ratio of molecular diffusivities in air for H$_{2}$O and HDO. The effect can be large at high supersaturations and cold temperatures: in upper tropospheric cirrus forming from sulphate aerosols ($S_i$=$1.5$, T=190 K), $\alpha_{\mathrm {eq}}$ would be reduced by 13\% even using a conservative estimate of d=1.0164 (\cite{Cappa03}, one of the lowest published estimates of d).
The diffusive model is widely used but poorly validated; neither of two published experiments demonstrating kinetic effects in ice growth showed quantitative agreement with Eq.\ (\ref{eq:alphak})  \cite{Jouzel84,Uemura05}.

More recently, studies have proposed modifying the diffusive model to include surface kinetic effects \cite{Nelson11}. Such effects would be especially important in the upper troposphere, where ice crystals are small.
(See Methods.) The ratio of deposition coefficients for H$_{2}$O and HDO (x) has never been measured, but suggested plausible values of 0.8-1.2 could modify the kinetic isotope effect (in our example, with 1 $\mu$m particles, from 13\% to 11-15\%). Neither previous experimental study of kinetic fractionation was sensitive to surface kinetic effects, since both involved large dendritic crystals.

To investigate both equilibrium and kinetic isotopic effects at low temperatures, we carried out a series of experiments at the Aerosol Interactions and Dynamics in the Atmosphere (AIDA) cloud chamber in the 2012-2013 IsoCloud campaign. We determine isotopic fractionation not from static conditions as in previous studies, but by measuring the evolving concentrations of HDO and H$_{2}$O vapour as ice forms. Results reported here are derived from measurements from a new \textit{in-situ} instrument (the Chicago Water Isotope Spectrometer, ChiWIS, see \cite{Sarkozy15}) and from AIDA instruments measuring total water, water vapour, temperature, and pressure. 

AIDA is a mature facility that has been widely used for studies of ice nucleation and cirrus formation (e.g.\ \cite{Moehler03,Cziczo13}); we employ the same procedure for adiabatic expansion and cooling as previous studies. The analysis here uses 28 experiments during the March-April 2013 campaign, 
covering a wide range of conditions: initial temperatures from 234--194 K, mean supersaturation over ice ($S_i$) of 1.0--1.35, mean ice particle diameter of 2--14 $\mu$m, and ice nucleation via mineral dust, organic aerosols, and sulphate aerosols. (See also S2 and Table S3.)
  Temperatures were set below the homogeneous freezing limit of $\sim$233 K to preclude coexistence of liquid and ice phases. 

\begin{figure}[h!]
\centering
 \includegraphics[width=0.45\textwidth]{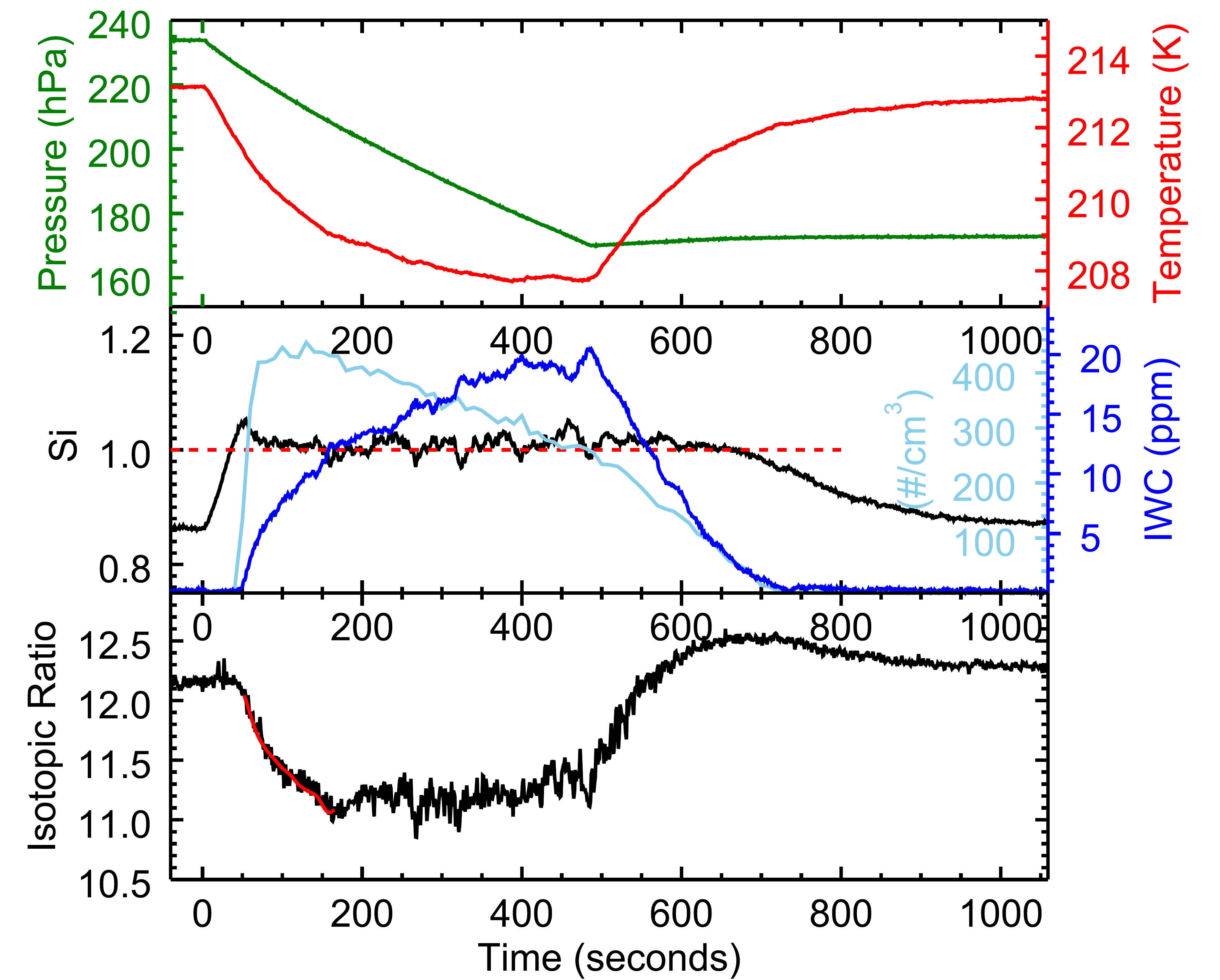}
\caption{Typical adiabatic expansion experiment. Pressure drop (top, green) causes drop in temperature (top, red) for $\sim$2 minutes before thermal flux from the wall becomes important. Ice formation (center: light blue, ice number density; dark blue, ice water content) begins when critical supersaturation (center, black) is reached. Vapour isotopic ratio (bottom, black, doped to $\sim$12x natural abundance) shows 3 stages: initial decline as ice growth draws down vapour; constant period when ice growth is driven by wall flux; and final rise as ice sublimates. Fractionation factor is derived from model fit to initial period (bottom, red). After sublimation, vapour isotopic ratio exceeds starting value because of wall contribution; system then reequilibrates over $\sim$5 minutes. Fluctuations while ice is present reflect inhomogeneities due to turbulent mixing.\label{expansion}}
\end{figure}

During a typical expansion experiment (Figure \ref{expansion}), chamber air cools by 5-9 K, 
causing nucleation and growth of ice particles \cite{Moehler06}. The isotopic ratio in the water vapour phase initially lightens as the heavier isotopologues preferentially deposit as ice. After several minutes, the walls (prepared with a thin ice layer) become a source of both water vapour and heat \cite{Cotton07}. Ambient supersaturation during ice growth depends on the nucleation threshold ($S_i\sim$1-1.2 for heterogeneous and 1.4-1.6 for homogeneous nucleation), on ice growth rate, and on ice particle number density. Most experiments reach saturation quickly, but in 
dilute conditions ice growth can take several minutes to draw chamber vapour down to equilibrium.

\begin{figure*}[ht!!!]
\centering
 \begin{tabular}{@{}cc@{}}
   \includegraphics[width=0.5\textwidth]{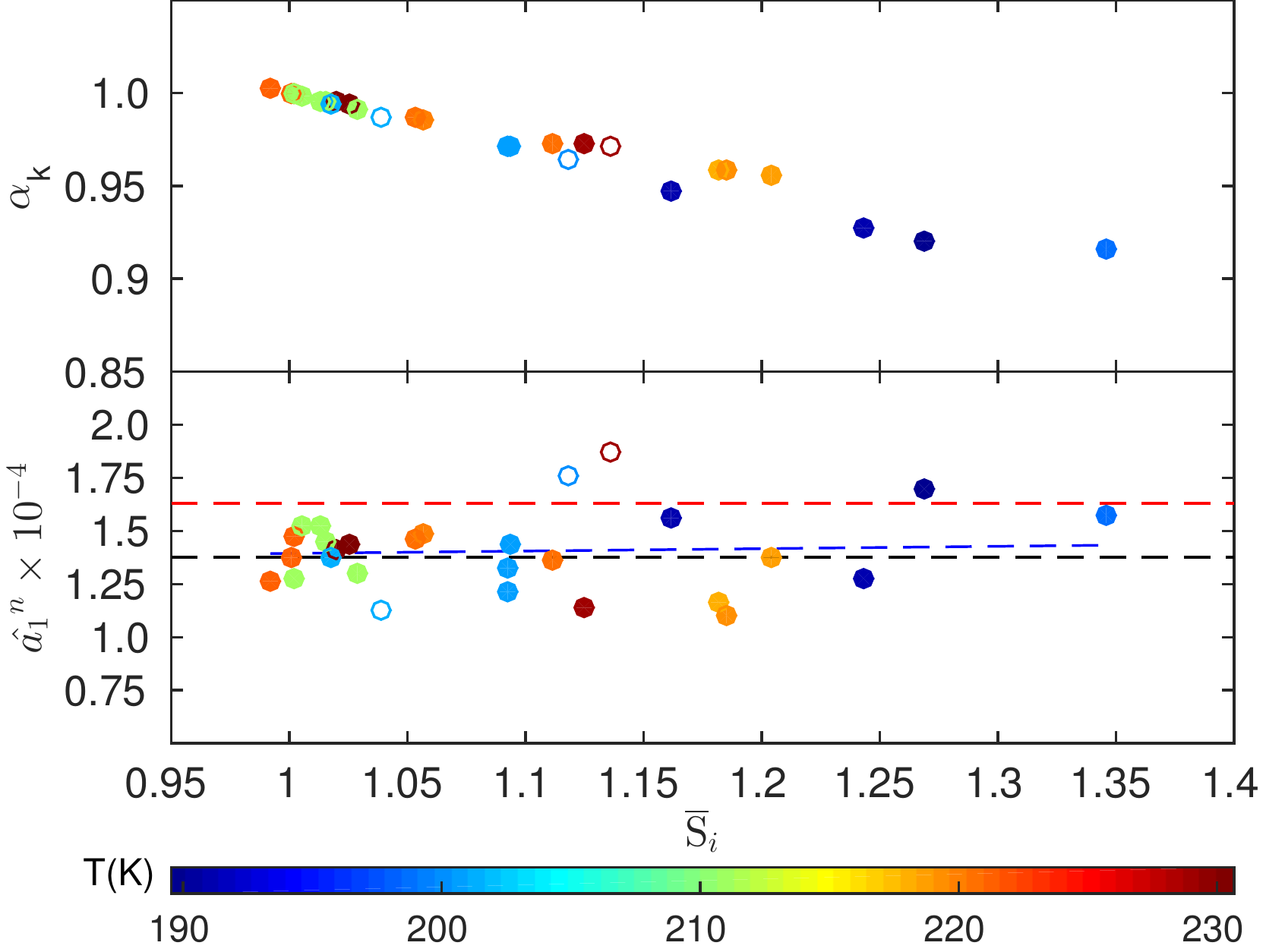} 
  \includegraphics[width=0.5\textwidth]{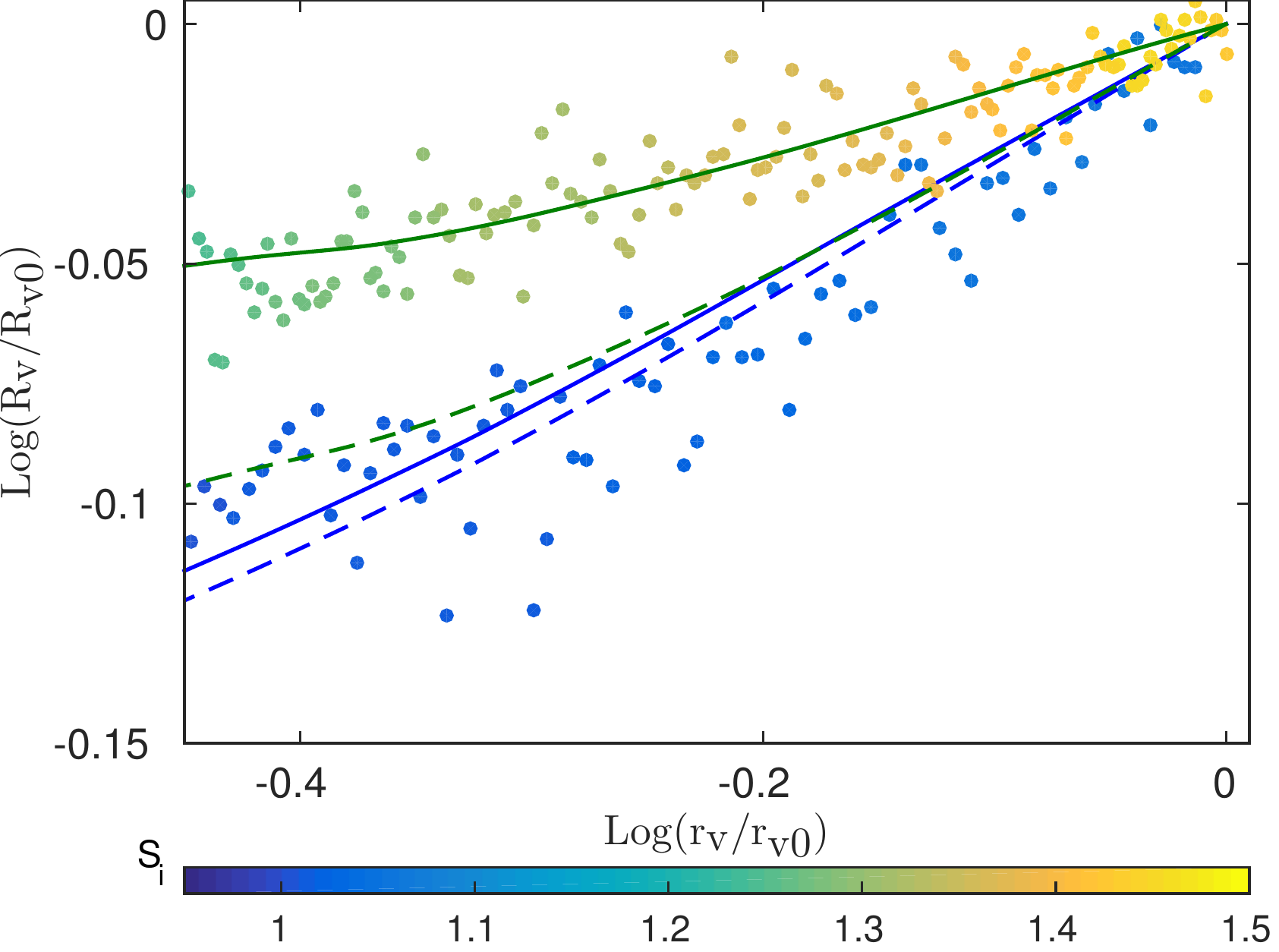}
 \end{tabular}
\caption{\textbf{Left:} Comparison of assumed kinetic and derived equilibrium isotope effects, to test the validity of diffusive model for kinetic isotope effects.  Top panel: assumed $\alpha_{\mathrm{k}}$ for each experiment using Eq.~(\ref{eq:alphak}) and d=1.0164, plotted against the deposition-rate-weighted average supersaturation. Bottom panel: the derived $\alpha_{\mathrm{eq}}$ for each experiment in a form that normalises temperature dependence, as point by point computed slopes $\hat{a_{1}}^{n}=(\log{\alpha_{n}/\alpha_{0}})/(1/T_{n}^{2}-1/T_{0}^{2})$. (See cartoon in Fig.\ S9.) Dashed lines show values corresponding to $\alpha_{\mathrm{eq}}$ in M67 (red) and in fit to all experiments in this work (black). Blue dashed line is a weighted fit through experiments (excluding outlier experiments \#4, 26, and 48, open circles; see Fig.\ S5 and discussion in S6). Data show negligible trend with supersaturation, suggesting that the diffusive model and value for d are approximately correct.  \textbf{Right:} Example of reduction in isotopic partitioning when ice grows in supersaturated conditions. Data points show 1-second measurements of $R_v$=[HDO]/[H$_2$O] in two expansion experiments (\#27 and 45) at similar temperatures but with differing $S_i$ (mean 1.01 and 1.35), plotted against evolving water mixing ratio $r_v$. (Both axes are scaled to initial values; only relative changes are physically meaningful.) The slope gives the effective fractionation $\alpha_{\mathrm{eff}}-1$. (Deviations from linearity result from wall flux and changing $S_i$.) The two experiments show different effective fractionation (solid lines) but similar derived equilibrium fractionation (dashed lines).  \label{kinetic}}
\end{figure*}

We derive the equilibrium fractionation factor for each expansion experiment by fitting a model that includes equilibrium fractionation modified by kinetic effects ($\alpha_{\mathrm {eff}}$=$\alpha_{\mathrm {eq}}\cdot\alpha_{\mathrm k}$) and by the isotopic contribution of any wall flux (See Methods and S3-S4.). In the absence of wall outgassing, vapour isotopic composition would evolve as simple Rayleigh distillation, with vapour progressively depleted as HDO is segregated into the ice phase \cite{Rayleigh02}. The evolving depletion then allows for the extraction of the equilibrium fractionation factor, given a model for the kinetic modification $\alpha_{\mathrm k}$. In the analysis here, we also account for deviations from Rayleigh distillation when the wall contribution becomes non-negligible. 

Experiments in supersaturated conditions show suppressed isotopic fractionation  (Fig.\ \ref{kinetic}, right) and allow us to quantitatively test models of kinetic isotope effects. Because equilibrium fractionation should depend only on temperature, any dependency of the retrieved $\alpha_{\mathrm{eq}}$ on supersaturation would imply an over- or under-correction for kinetic effects. When we estimate $\alpha_k$ with the standard diffusive model of Eq.~(\ref{eq:alphak}) and d=1.0164 \cite{Cappa03}, the resulting 
fitted values for $\alpha_{\mathrm{eq}}$ show negligible dependence on supersaturation (Fig.\ \ref{kinetic}, left). That is, the diffusive model and a standard assumption for d explain results across a wide range of $S_i$. 

We use this test of retrieving a consistent $\alpha_{\mathrm{eq}}$ independent of $S_i$ to place rough constraints on the isotopic diffusivity and deposition coefficient ratios (d and x). While d and x cannot be constrained simultaneously, each can be estimated given an 
assumption about the other, along with 1$\sigma$ bounds from propagation of uncertainties. A pure diffusive model yields an optimal d slightly below all published estimates, though with an upper bound encompassing the highest literature value: 1.009 $\pm 0.036$, while published estimates of d evaluated at 190 K span 1.015--1.045. (See S6 and Fig.\ S10; a surface kinetic model with x=1 is essentially identical.)  For a model with surface kinetic effects, we obtain x slightly below 1 regardless of the assumed diffusivity ratio: for example, x=0.957 and 0.924 (both $\pm 0.22$) for d=1.0164 \cite{Cappa03} and 1.0251 \cite{Merlivat78}. The bounds on this estimate are consistent with the plausible range x=0.8-1.2 suggested by \cite{Nelson11}. (See S6 and Fig.\ S11). 

\begin{figure}[ht!!!]
\centering
\includegraphics[width=0.5\textwidth]{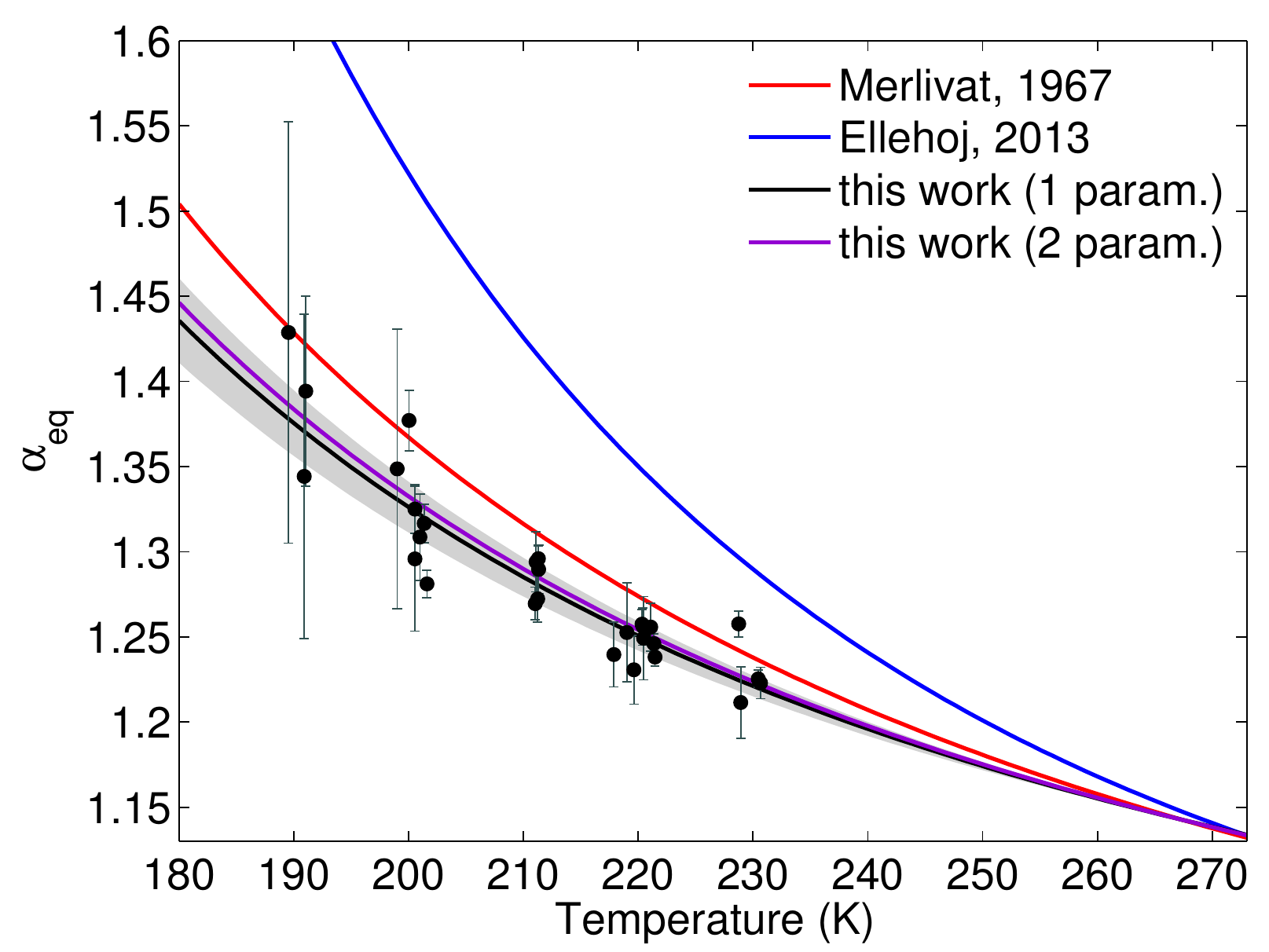}
\caption{Equilibrium vapour-ice fractionation factor for HDO/H$_{2}$O ($\alpha_{\mathrm {eq}}$) derived from 28 individual IsoCloud experiments. Black and purple lines show global fits through all experiments for two data treatments (black: 1-parameter fit, wall flux composition $R_w$ assumed to be that of ice initially at equilibrium with chamber vapour; purple: 2-parameter fit, $R_2$ as independent parameter).  Dots show individual experiments (1-parameter), and grey shading the 3$\sigma$ confidence interval. Error bars represent 2$\sigma$ uncertainties in fits to individual experiments. (These underestimate experimental error at warmer temperatures; see S4.) Solid lines show M67 (red) and E13 (dark blue). (See Methods, S4, and S5.) Results robustly imply weaker temperature dependence of $\alpha_{\mathrm{eq}}$ than in M67. \label{fig:equilibrium}}
\end{figure}

Finally, we determine the temperature dependence of the equilibrium fractionation factor by taking a weighted global fit of all 28 individual experimental values for $\alpha_{\mathrm {eq}}$ (assuming the $1/T^2$ functional form of Eq.\ (\ref{eq:tempdepend})). The resulting temperature dependence lies far below E13, and slightly below M67 (Fig.\ \ref{fig:equilibrium}).  The distinction from M67 is significant to a 3$\sigma$ confidence interval and robust to assumptions made in fitting and in modelling kinetic isotope effects. (See Methods, S5, and S6). In the two fitting methods shown in Figure \ref{fig:equilibrium},  global estimates for $\alpha_{\mathrm{eq}}$ differ by $< 10^{-2}$ throughout the experimental temperature range. Derived constants for our 1-parameter fit (see Methods) are $a_{0}=-0.0559$ and $a_{1}=13525$. (Compare to M67: $a_{0_{M67}}=-0.0945$ and $a_{1_{M67}}=16289$.) 

This study represents the first direct measurement of the equilibrium fractionation factor between HDO and H$_{2}$O at the cold temperatures required for cirrus cloud formation, and 
the first quantitative confirmation of models of kinetic isotope effects in ice deposition. 
Our results imply a slightly weaker temperature dependence for $\alpha_{\mathrm{eq}}$ and therefore lower equilibrium fractionation than assumed by M67, and 
rule out the substantial upward revision proposed by E13.
Experiments in supersaturated conditions imply that kinetic fractionation is well-characterized by published models. While sensitivity tests provide only broad constraints, they are suggestive of slightly weaker kinetic isotope effects than obtained with typical parameter values. These results may imply diffusivity ratios at the lower end of literature values and/or that HDO has a slightly higher probability of incorporation into the ice matrix than H$_{2}$O.
These conclusions are limited by the small number of IsoCloud experiments;  a larger sample size in supersaturated conditions would provide a more rigorous test. 
The paucity of previous measurements of fundamental properties of stable water isotopologues at cold temperatures highlights the need for additional research, 
since accurate understanding of isotopic effects during ice deposition is 
needed for interpreting isotopic water measurements for atmospheric and climate science.
The IsoCloud results demonstrate that chamber-based measurements, which provide the most realistic laboratory simulations of ice growth in cirrus clouds, can be an important tool for this purpose. 

\small
\section*{Methods}
\textbf{Experiments}.
Each campaign day involved 4--6 expansion experiments at the same initial $T$, separated by 1--2 hours to re-establish equilibrium. To boost signal to noise, all water introduced into AIDA was isotopically doped to produce HDO/H$_{2}$O ratios of $\sim$10-20$\times$ natural abundance (VSMOW). In most experiments, chamber walls were prepared with a thin layer of isotopically doped ice, which is in isotopic equilibrium with the chamber vapour before experiments begin. For comparison, one experimental day involved dry walls; results with both treatments are consistent.   
(Table S3 and Fig.\ S4 show all experiments.) 
HDO and H$_{2}$O are measured with the ChiWIS tunable diode laser absorption spectrometer \cite{Sarkozy15},  and 
 ice content is determined from ChiWIS water and total water from the APeT instrument \cite{Lauer07,Skrotzkithesis}. (Wall outgassing flux is determined from changes in total water).
An optical particle counter measures ice crystal number, and size is then estimated by assuming spherical particles. 
In cases of thick ice clouds,
slight corrections for backscatter effects in ChiWIS are determined by comparison to water vapour from the SP-APicT instrument \cite{Skrotzkithesis}.  See S2 for further information about instruments, experiments, data treatment, and campaign. 

\textbf{Isotopic modelling}.
Because most ice growth occurs in the first few minutes of each experiment and wall contribution grows over time, we fit only the initial part of each experiment when ice deposition dominates (54-223 seconds, see S4.1 for selection criteria). 
 We fit each experiment using a model derived from mass balance over H$_{2}$O and HDO : 
\begin{equation}
\label{masterode}
\frac{dR_v}{dt} = -\left(\alpha_{\mathrm{eff}}-1\right) R_v \frac{P_{\mathrm{vi}}}{r_v} + \left(\gamma-1\right) R_v \frac{S_{\mathrm{wv}}}{r_v}. \label{eq:model}
\end{equation}
where $\alpha_{\mathrm {eff}} = \alpha_{\mathrm {eq}}\cdot\alpha_{\mathrm k}$; $\gamma\equiv R_w/R_v$, the ratio of isotopic composition of wall flux to that of bulk vapour; and $P_{\mathrm{vi}}$ 
and $S_{\mathrm{wv}}$ 
represent the loss of vapour to ice formation and the source of vapour from wall outgassing. (See also S3.) We estimate the isotopic composition of the wall flux either by fitting $\alpha_{\mathrm{eff}}$ and $\gamma$ independently (2 parameter fit) or by assuming that outgassing is non-fractionating sublimation of ice that had equilibrated with chamber vapour, i.e.\ $R_w = \alpha_{\mathrm{eq,0}}\cdot R_{v0}$  (1 parameter fit). We derive uncertainty estimates for global fitting, shown in Figure \ref{fig:equilibrium}, from the 2-parameter case. 
(See S4 for fitting of individual experiments and uncertainty treatment and S5 for global fitting.)

For estimating kinetic isotopic effects, we assume the Murphy-Koop parameterization for saturation vapour pressure $S_{i}$ \cite{Murphy05}. 
When incorporating surface kinetic effects following \cite{Nelson11}, the diffusivity ratio $\mathrm{d}$ in Eq.\ (\ref{eq:alphak}) is replaced by $(\mathrm{d}k+\mathrm{x}\mathrm{y})/(1+k)$, where x is the ratio of deposition coefficients, y is the ratio of thermal velocities ($\sqrt{19/18}$), and the dimensionless coefficient $k\equiv rv\beta/4D_{v}$, where $r$ is the ice particle radius and $v$, $D_{v}$, and $\beta$ are the thermal velocity, diffusivity in air, and deposition coefficient for H$_2$O, respectively. Values for $k$ in experiments (2-15) follow ice particle diameter (2-14 $\mu$m, $>$5 $\mu$m for $T> 215$ K and smaller at lower temperatures). \onecolumn

\vspace{10mm}
\noindent
Funding for this work was provided by the National Science Foundation (NSF) and the Deutsche Forschungsgemeinschaft (DFG) through an International Collaboration in Chemistry grant (CHEM1026830). K.L.\ acknowledges support from a National Defense Science and Engineering Graduate Fellowship and an NSF Graduate Research Fellowship, and L.S.\ from a Camille and Henry Dreyfus Postdoctoral Fellowship in Environmental Chemistry. The many IsoCloud participants contributed greatly to the success of the campaign, including Stephanie Aho, Jan Habig, Naruki Hiranuma, Erik Kerstel, Benjamin K\"uhnreich, Janek Landsberg, Eric Stutz, Steven Wagner, and the AIDA technical staff and support team.  We thank Marc Blanchard, Won Chang, Albert Colman, Nicolas Dauphas, and William Leeds for useful discussions and comments. 

\clearpage


\section*{Supplementary material}

# Contents

# S1 Review of previous determinations of $\alpha_{\text{eq}}$

A total of eight previous works have attempted to determine the fractionation factor for vapour over ice for the HDO to $H_2O$ system, five experimental measurements and three theoretical calculations. See Table S1 for summary and Figure S1 for plot.

The measurements can be roughly categorized as "indirect" or "direct" measurements. In the three indirect measurements, the quantities measured are typically the vapour pressures over $D_2O$ and $H_2O$ ice, and the fractionation factor is derived by assuming the law of geometric means [1, 2]. In the two previous direct measurements, fractionation between phases is determined by measuring the isotopic ratio in a vapour stream at two points, before and after vapour comes into contact with an ice surface and presumably reaches equilibrium with it. Theoretical values are derived from *ab initio* calculations. Because of the difficulty of determining the crystal partition function, these are not expected to provide tight constraints on the fractionation factor.

| Measurement | Temp. Range | Type | Method |
|---|---|---|---|
| Matsuo, et al., 1964 [3] | 235 - 273 K | Experimental (indirect) | Vapour pressure, $D_2O$ and $H_2O$ |
| Merlivat and Nief, 1967 [4] | 233 - 273 K | Experimental (direct) | Mass spectrometry |
| Van Hook, 1968 [5] | 233 - 273 K | Theoretical | Effective force field |
| Johansson et al. 1969 [6] | 253 - 273 K | Experimental (indirect) | Specific heat, latent heat |
| Pupezin et al., 1972 [7] | 209 - 273 K | Experimental (indirect) | Vapour pressure, $D_2O$ and $H_2O$ |
| Méheut et al., 2007 [8] | 203 - 273 K | Theoretical | Density functional theory |
| Ellehøj et al., 2013 [9] | 223 - 269 K | Experimental (direct) | Mass spectrometry |
| Pinilla et al., 2014 [10] | 190 - 273 K | Theoretical | Path integral molecular dynamics |

Table S1: Previous measurements and theoretical determinations of the equilibrium fractionation factor for vapour over ice for the HDO/$H_2O$ system.

# S2 IsoCloud campaign and instruments

The IsoCloud campaigns at the AIDA Aerosol and Cloud Chamber ran from Spring 2012 to Spring 2013, with the first two campaigns dedicated to integration and engineering tests and the final two to science. Instruments participating in these campaigns were provided by research groups from several institutions. Details of instruments used in the analysis are given in S2.1 and instrument placement is shown in Figure S2. All data shown in this work is taken from IsoCloud 4 (March 2013), since low isotopic doping during IsoCloud 3 (October 2012) limited data utility.

The main goals of the IsoCloud campaigns were to

- Develop and test isotopic water vapour instruments in a controlled laboratory setting.
- Measure the equilibrium fractionation factor $\alpha_{\text{eq}}$ between vapour and ice in atmospheric conditions typical of the upper troposphere and lower stratosphere.
- Verify kinetic fractionation effects due to diffusion limitation at low temperatures.
- Investigate potential surface inhibition effects that could lead to anomalous supersaturation and could potentially have isotopic implications.
- Investigate potential metastable phases of ice that could form at low temperatures, and could potentially have isotopic implications.

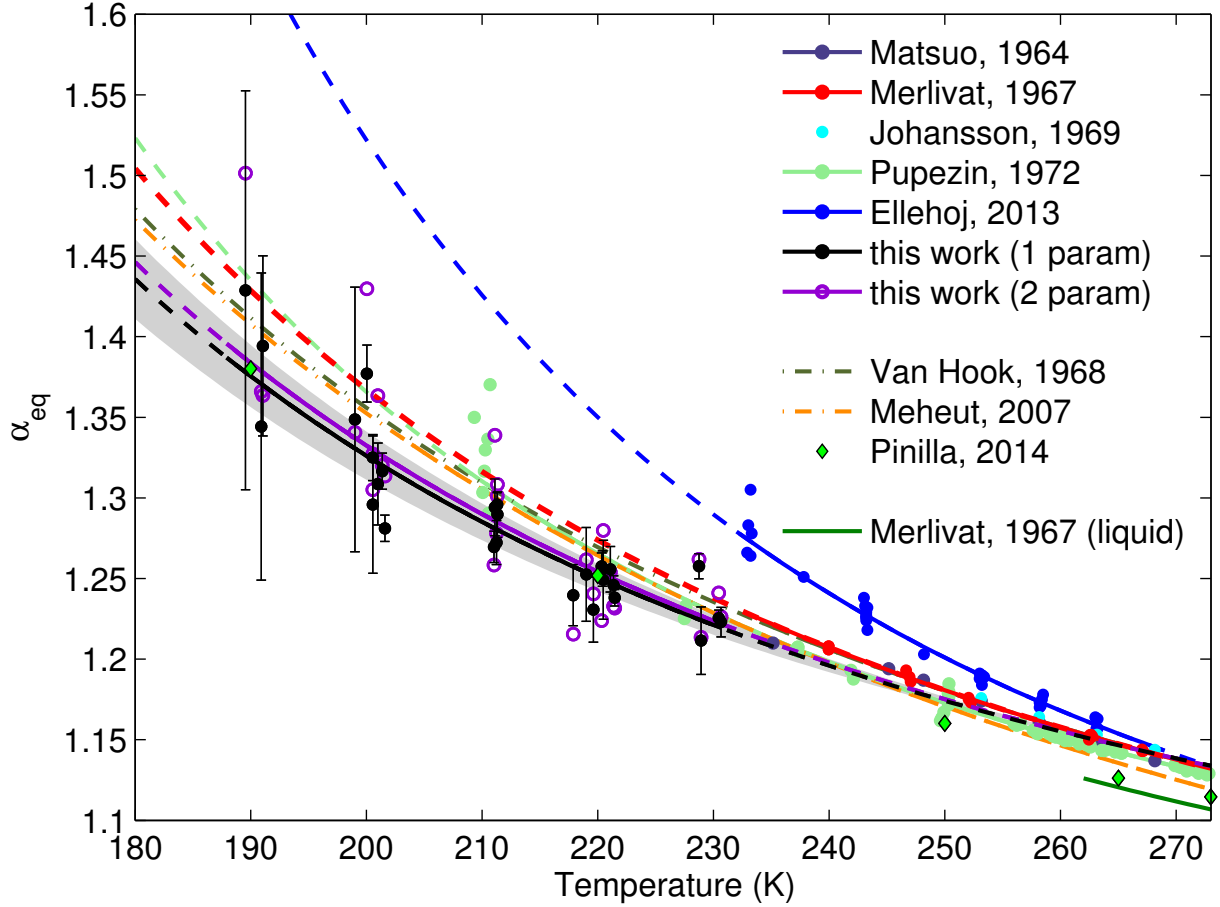

Figure S1: All previous determinations of the fractionation factor between vapour and ice for $HDO/H_2O$, and new measurements from this work (shown for two fitting cases). For all measurements, dots are experimental values, solid lines are parameterizations derived from those measurements in the experimental temperature range, and dashed lines are extrapolations outside the experimental range. Calculations and parameterizations from theory are shown in diamonds and dashed-dotted lines. (Theoretical calculations are not expected to provide tight constraints on the fractionation factor.) The fractionation factor for the vapour-liquid transition is shown for reference; preferential partitioning in the vapour-liquid system should be less than in the vapour-solid system. Error bars shown for this work are $2\sigma$ uncertainties, shown for clarity only on the one-parameter case. We assume identical uncertainties for both cases; see S4-S5 for determination and discussion. These uncertainties are used as weights in fitting for the global temperature dependence of $\alpha_{eq}$.

## S2.1 Instruments used in isotopic analysis

The isotopic ratio measurements used in the analysis are taken from the ChiWIS instrument, but the fractionation factor analysis requires data from water measurements by two other instruments as well and from facility temperature and pressure sensors. Measurements used are summarized in Table S2 and discussed in more detail below.

| Exp. Observable | Instrument | Technique |
|---|---|---|
| $T_{gas}$ | AIDA sensors | thermocouples |
| $p_{gas}$ | AIDA sensors | – |
| HDO/$H_2O$ vapour ratio | ChiWIS | in-situ, TDL multi-pass |
| $H_2O$ vapour mixing ratio | ChiWIS | in-situ, TDL multi-pass |
| $H_2O$ vapour mixing ratio | SP-APicT | in-situ, TDL single-pass |
| Total $H_2O$ (vapour + ice) | APeT | extractive, TDL multi-pass |
| Ice number density | WELAS | optical particle counter |

Table S2: Experimental observables used to determine $\alpha_{\text{eq}}$ during the IsoCloud campaigns. All water measurements are made by tunable diode laser (TDL) absorption spectroscopy. APeT and SP-APicT are AIDA instruments that measure $H_2^{16}O$, and ChiWIS is an isotopic water vapour instrument integrated into AIDA specifically for HDO and $H_2^{16}O$ measurements during IsoCloud.

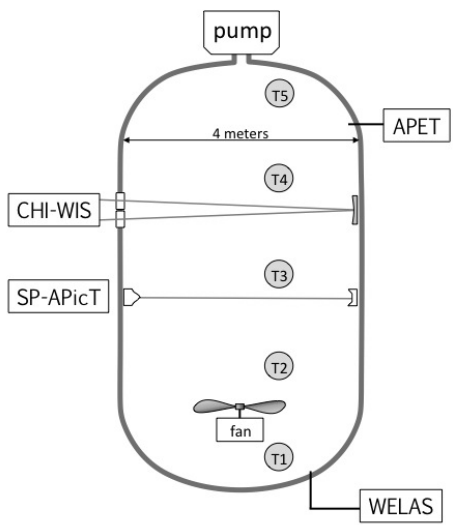

Figure S2: Positioning of the instruments used in this analysis during the IsoCloud experiment campaigns. (Additional instruments also participated in the IsoCloud campaigns.) We take gas temperature as the average of thermocouples T1-4. Data from the WELAS optical particle counter are used to derive the effective ice particle diameter listed in Table S3 and in calculating kinetic isotope effects.

### S2.1.1 AIDA gas temperature and pressure measurements

Gas temperature inside the containment vessel is measured by five thermocouples at different heights along an axis about 1 m from the center of the vessel. (See Figure S2.) We take as the gas temperature the average of the four lowest thermocouples (T1-4); the top of the chamber is less well mixed and experiences fluctuations that are not characteristic of the bulk air being sampled. Mixing is provided by a fan located 1 m above the floor of the aerosol chamber. Individual thermocouples show variations of ±0.3 K during expansion experiments

and ±0.15 K during stable periods between pumpdowns, on timescales of 30 s, likely due to turbulent mixing [11]. These inhomogeneities form an additional source of uncertainty for experiments. Gas temperatures and pressures are used in calculation of spectral line shapes, saturation vapour pressure, and in the conversion of number density to mixing ratio.

### S2.1.2 APeT total water measurement

APeT (AIDA PCI extractive TDL) is a tunable diode laser absorption spectrometer permanently installed at the facility. It measures total water (ice and vapour) by extracting gas from the chamber through a heated inlet [12, 13]. We assume a 17 second delay in the APeT extractive measurements based on previous comparison of *in situ* and extractive instruments [13]. This delay is consistent with the timing of observed changes in isotopic ratio during IsoCloud pumpdowns. We use total water to determine the rate of deposition of vapour to ice, the contribution of vapour from the chamber walls, and the total ice content.

### S2.1.3 SP-APicT water vapour measurement

SP-APicT (single pass AIDA PCI in cloud TDL) is a tunable diode laser absorption spectrometer installed at AIDA that measures *in situ* vapour $H_2O$ with a single-pass configuration (4.1 m pathlength) [13]. Because the SP-APicT optical path involves no reflections, the measurement is not affected by backscattering from chamber ice particles. Water vapour measurements from ChiWIS to SP-APicT show a consistent ratio of ∼1.025 in no-cloud conditions at temperatures above 205 K, likely the result of systematic error in the linestrengths used in retrievals for one or both instruments [14, 15]. (At lower temperatures, water content is below the SP-APicT dynamic range.) SP-APicT is used for backscattering corrections to ChiWIS water vapour and vapour isotopic ratio in experiments with dense ice clouds.

### S2.1.4 WELAS particle counters

The WELAS instruments are optical particle counters that can measure ice particle number concentrations for particles in a specified size range. Two instruments were used during the IsoCloud campaigns, with effective spherical size ranges of 0.3–46 and 4–237 $\mu$m, time resolution of 5 s, and accuracy estimated at ±20 % [16]. (Error is driven by any non-spherical ice crystal morphology.) In this analysis, WELAS data are used to approximate effective ice crystal radii for use in the kinetic fractionation model that includes surface kinetic effects.

### S2.1.5 ChiWIS $H_2O$ and HDO measurements

The Chicago Water Isotope Spectrometer (ChiWIS) is a tunable diode laser absorption spectrometer that scans across both HDO and $H_2O$ spectral lines, allowing for a simultaneous retrieval of the concentration of both isotopologues in the vapour phase inside AIDA. The spectrometer is used in a non-resonant multi-pass White Cell configuration (set between 196.3–256.5 m) inside the cloud chamber, allowing for *in-situ* measurements of the evolving isotopic composition inside AIDA [17, 18]. The instrument design, data acquisition, fitting, and performance during the IsoCloud campaigns are described in detail in [19].

The dynamic range of ChiWIS for isotopic measurements in IsoCloud conditions is ∼0.5–400 ppm water at pressures ∼100–300 mb (producing 0.06–39% absorption for $H_2O$ and 0.03–19% for HDO doped 15x). Measurements of isotopic composition are limited at high temperatures by saturation of the $H_2O$ line and at low temperatures by signal-to-noise on the HDO line. (The low-temperature limit could be modified by higher isotopic doping levels.)

We apply a calibration to ChiWIS measurements during some IsoCloud experiments with ice clouds dense enough that backscattered light onto the detector produces artifacts. The dominant effect of dense ice clouds is to reduce total laser power reaching the ChiWIS detector (by up to 95%), but a secondary effect is that some light scatters back onto the detector after traversing a shorter distance than the intended optical pathlength. That backscattered contribution reduces all apparent line depths. If laser power at the position of HDO and $H_2O$ lines were identical, any modification would be identical for both species and would not affect their ratio. In actuality, the $H_2O$ and HDO lines are affected slightly differently. Since SP-APicT is not affected by backscattering, we identify the experiments that may experience artifacts by comparing measured concentrations of $H_2O$ by SP-APicT and ChiWIS. When needed (17 of 28 experiments), we use a simple model to reconstruct ChiWIS spectra without contributions from backscattered light. Under the harshest experimentally realized conditions, the applied correction to the isotopic ratio is 0.22% out of a total ratio decline of 13.2% (at 225 K; see Table S3). (The adjustment on water vapour alone is larger, with maximum value 11%.) At lower temperatures, water vapour content in the chamber is too small to produce clouds thick enough to cause deviations in absorption spectra.

Our backscatter correction model assumes that backscattered light has a path length so short that it can be approximated as an absorption-free offset to the raw data. For each spectrum recorded, we calculate the effect on the isotopic ratio using synthetic spectra. First, we verify that the ratio between ChiWIS and SP-APicT is identical before and after the presence of ice particles. We then fit the raw spectral data to a model that assumes the true concentration is given by SP-APicT scaled up by this ratio, and that the spectrum of the main beam sits atop a frequency-independent offset introduced by backscatter into the detector. The size of this offset is the only free parameter in this fit. We then remove the calculated offset from the ChiWIS data and re-fit for the HDO concentration.

The robustness of the procedure is demonstrated by the relative consistency of inferred fractionation factor values at a given temperature. Because ice cloud properties vary between pumpdowns, each temperature cluster of experiments involves different degrees of influence from backscatter. We see no evidence of systematics in the final calibrated data.

## S2.2 Preparation of AIDA for isotopic measurements

AIDA was chosen as a setting for an isotopic water measurement campaign because of its extensive history of use for cirrus experiments, its well-characterized facility instruments, and its large volume (84 $m^3$), which mediates wall effects. On each of nine experimental days (4-6 individual pumpdowns) the AIDA chamber wall temperature remained approximately constant. Between experimental days, chamber temperature was adjusted at night (typical rates ∼4 K/h) to a new setpoint. To test for any systematics, some temperatures were repeated on multiple days, so the total campaign comprises 6 temperature groupings, 3 of which were repeated. During IsoCloud 4, chamber preparation followed standard procedure

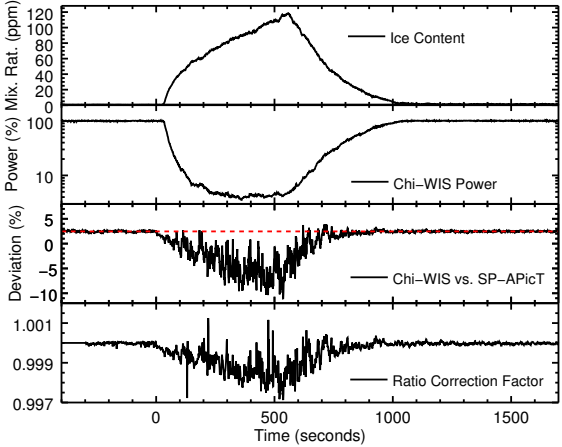

Figure S3: Example of a pumpdown in thick ice cloud conditions (Experiment #8; the most extreme case). Time 0 marks the start of pumping; ice formation begins ∼20 seconds after and grows rapidly, attenuating the collected ChiWIS power, with peak loss ∼95%. Power recovers as the ice cloud dissipates, indicating that no change in instrument alignment has occurred. ChiWIS $H_2O$ is typically 2.5% above that in SP-APicT; as ice content grows, ChiWIS drops relative to SP-APicT. The applied isotopic ratio correction is much smaller than that for water vapour (max 0.22%) since HDO and $H_2O$ are affected similarly. Ice content is determined using APeT total water.

with some special adaptations for isotopic water measurements, described in more detail below. During most of the IsoCloud 4 campaign, the chamber was prepared with ice-covered walls to ensure a known isotopic composition of flux from the walls. Both wall ice and water vapour in the cloud chamber were isotopically doped above natural abundance to provide a larger signal for isotopic measurements. On one experimental day, all pumpdowns were conducted with dry walls.

*Chamber preparation.* At the end of each experiment day, the chamber is purged with a pump-and-flush cycle: eight times first pumping out to 1 hPa, then filling with synthetic dry air at 10 hPa. The chamber is then pumped out completely to a pressure of ∼0.01 hPa to remove all aerosols. (Background aerosol concentrations are typically < 0.1 particle per $cm^{-3}$.) In the morning of the next experimental day, the desired amount of water vapour for the next experiment is added to the chamber by opening a valve leading (via a heated stainless steel tube) to a heated water reservoir containing nanopure-quality water with isotopic composition enriched in HDO by approximately ×10-20 and in $H_2^{18}O$ by approximately ×2 compared to natural abundance (VSMOW) [20, 21, 22]. (HDO enrichment is achieved by adding $D_2O$, which then partitions statistically.) After water addition, the chamber is filled with dry synthetic air ($N_2$ and $O_2$ only) up to the desired pressure for the first experiment. Most experimental days in IsoCloud 4 begin with a "reference pumpdown" with essentially no ice nuclei present, so that no condensation occurs. Aerosols are then added to ensure formation of ice crystals in all subsequent pumpdowns.

*Ice-covered wall preparation.* On most days, sufficient water is injected into the chamber to not only saturate chamber air but also to form a thin coating of ice on chamber walls. (The wall temperatures are typically ∼ 0.3 to 0.9 K lower than chamber air temperature). If the ice coating were uniform, the total amount would correspond to a layer ∼2 μm thick on the 110 $m^2$ chamber surface. In practice, non-uniform wall temperatures may produce a inhomogeneous ice layer [23]. The ice layer serves as a water source that recharges the chamber following each pumpdown. When the chamber is static, the walls maintain the chamber air at between 80-85% of saturation vapour pressure, since wall temperatures are slightly lower than chamber air temperature. After a pumpdown, once ice particles in the

chamber have fully sublimated, chamber water vapour and the isotopic ratio of the gas re-equilibrates with the walls on a timescale of ∼5 minutes. In our data analysis, fits of the isotopic composition of the wall outgassing component are consistent with the isotopic signature of ice that had been in equilibrium with chamber vapour before the pumpdown.

*Dry wall preparation.* As a test, one experimental day (experiments #39-43) was conducted with bare aluminum walls. On this day, operators added water to the chamber in successive steps until chamber air reached just below saturation vapour pressure at the wall temperature. The chamber was then left static for ∼30 minutes to allow chamber vapour and walls to equilibrate. After each pumpdown (with attendant loss of water), water was again added to bring the chamber close to saturation. Dry-wall pumpdowns feature a lower proportion of chamber vapour depositing as ice and smaller changes in vapour isotopic composition.

## S2.3  Summary of IsoCloud 4 experiments

The analysis here involves all 28 of the 48 total expansion experiments during IsoCloud 4 that produced useable isotopic ratio measurements during ice cloud formation. These data were taken on seven different experimental days. (See Table S3 for complete list and Figure S4 for isotopic evolution in individual experiments.)

Of the twenty experiments not analyzed, nine involved insufficient ice deposition: six intentional "reference" pumpdowns with no aerosol or mineral dust addition (#12, 18, 28, 34, 39, and 44), and three cases of unintentional insufficient addition (#2, 19, and 40). (We analyze only experiments in which at least 20% of the initial vapour deposits as ice.) On one experimental day, anomalously high noise on ChiWIS isotopic ratios precluded use of all experiments (#29-33). On the coldest experimental day (189 K), isotopic doping was insufficient for use of HDO measurements, precluding use of experiment #35; for the three following experiments (#36-38), the laser was tuned over a different spectral region to focus on $H_2O$, foregoing HDO measurements. ChiWIS did not record data during one experiment (#23), and another (#24) began before the ChiWIS laser was stabilized in temperature.

The analysis includes 3 experiments (#41-43) conducted on March 21, the day that the AIDA chamber was prepared with dry walls rather than with an isotopically doped ice layer. Isotopic retrievals from these experiments meet goodness-of-fit criteria but demonstrate significantly higher sensitivity to the choice of region length than experiments with ice-covered walls, including the set of experiments conducted at similar temperature (#6-11, on March 13). That sensitivity is reflected in larger uncertainty in the retrieval of the fractionation factor (Figure S5, top panel.) The isotopic signature of water vapour desorbing from the walls is also not likely to be exactly the value expected for equilibrated ice, implying some bias in the 1-parameter fit assumptions (See S4). In the 1-parameter fit, the dry-wall experiments fit systematically slightly low (Figure S5, bottom panel).

The complete dataset analyzed covers a range of pumpdown start temperatures from 194 to 234 K, producing water vapour mixing ratios between 2 and 380 ppmv. Maximum supersaturations ranged between 1.03 and 1.62, with the highest values in homogeneous nucleation experiments. High-supersaturation experiments are distributed across the temperature range. Our equilibrium fractionation factor retrievals show no systematic dependence on $S_i$. (See manuscript Figure 3 and Figures S10 and S11.)

| ID | $T_0$(K) | $\Delta T$(K) | $p_0$ (hPa) | $\Delta p$ (hPa) | $w_{eff}$ (cm/s) | $R_{vD,0}$ | $\delta R_{vD,0}$ (%) | $r_{v,0}$ (ppmv) | $\delta r_{v,0}$ (%) | $S_{max}$ | $\bar{S}_i$ | $d_{avg}$ ($\mu m$) | IN | $R_{vD,dev}$ (%) | $r_{dev}$ (%) |
|---|---|---|---|---|---|---|---|---|---|---|---|---|---|---|---|
| 1 | 234 | 7.8 | 299 | 65 | -370 | 16.5 | 8.5 | 380 | 39.0 | 1.21 | 1.12 | 14 | ATD | -0.18 | 5.4 |
| 2 | 233 | 6.5 | 300 | 100 | -130 | 16.6 | 4.0 | 366 | 17.7 | 1.24 | – | 7 | ATD | -0.06 | 1.5 |
| 3 | 233 | 6.4 | 300 | 101 | -120 | 17.1 | 6.0 | 377 | 28.6 | 1.03 | 1.02 | 10 | ATD | -0.06 | 1.4 |
| **4** | **233** | **9.1** | **300** | **131** | **-130** | **17.4** | **8.7** | **375** | **38.2** | **1.21** | **1.14** | **10** | **ATD** | **-0.09** | **2.4** |
| **5** | **233** | **9.1** | **300** | **132** | **-180** | **17.9** | **9.9** | **387** | **43.8** | **1.05** | **1.03** | **11** | **ATD** | **-0.09** | **2.3** |
| **6** | **223** | **6.6** | **300** | **71** | **-170** | **13.1** | **6.7** | **113** | **29.0** | **1.27** | **1.20** | **11** | **ATD** | **–** | **–** |
| **7** | **223** | **6.4** | **234** | **64** | **-140** | **12.7** | **10.6** | **147** | **35.3** | **1.03** | **1.00** | **6** | **ATD** | **-0.17** | **6.3** |
| **8** | **223** | **8.7** | **300** | **131** | **-200** | **12.8** | **13.2** | **114** | **46.4** | **1.04** | **1.00** | **8** | **ATD** | **-0.22** | **11.4** |
| 9 | 223 | 6.0 | 300 | 71 | -160 | 13.1 | 8.8 | 114 | 30.7 | 1.12 | 1.06 | 6 | ATD | -0.04 | 2.2 |
| **10** | **223** | **5.5** | **231** | **62** | **-130** | **13.1** | **7.7** | **147** | **29.7** | **1.10** | **1.05** | **6** | **ATD** | **-0.16** | **2.9** |
| **11** | **223** | **8.9** | **300** | **150** | **-180** | **13.2** | **14.7** | **115** | **47.3** | **1.03** | **0.99** | **7** | **ATD** | **-0.09** | **5.5** |
| 12 | 213 | 5.4 | 298 | 69 | – | – | – | – | – | – | – | – | – | – | – |
| **13** | **213** | **5.3** | **234** | **64** | **-130** | **11.5** | **9.6** | **40.6** | **33.1** | **1.06** | **1.03** | **2** | **ATD** | **-0.01** | **1.8** |
| **14** | **213** | **8.4** | **300** | **137** | **-160** | **11.9** | **14.4** | **30.9** | **46.8** | **1.04** | **1.00** | **2** | **ATD** | **-0.01** | **2.1** |
| **15** | **213** | **5.6** | **300** | **71** | **-160** | **12.0** | **9.7** | **31.3** | **63.4** | **1.04** | **1.01** | **2** | **ATD** | **-0.01** | **2.0** |
| **16** | **213** | **5.4** | **234** | **64** | **-140** | **12.1** | **9.5** | **39.9** | **32.1** | **1.03** | **1.02** | **2** | **ATD** | **-0.02** | **2.0** |
| **17** | **213** | **8.4** | **300** | **130** | **-150** | **12.2** | **15.6** | **31.1** | **48.3** | **1.04** | **1.01** | **2** | **ATD** | **-0.02** | **3.0** |
| 18 | 194 | – | – | – | – | – | – | – | – | – | – | – | – | – | – |
| 19 | 194 | 5.2 | 300 | 71 | -120 | 10.9 | 1.4 | 1.78 | 2.2 | 1.87 | – | 9 | ATD | – | – |
| **20** | **194** | **4.8** | **239** | **70** | **-90** | **10.5** | **12.7** | **2.13** | **36.2** | **1.46** | **1.16** | **2** | **ATD** | **–** | **–** |
| **21** | **194** | **7.6** | **300** | **131** | **-120** | **10.2** | **18.4** | **1.70** | **53.9** | **1.60** | **1.24** | **1** | **ATD** | **–** | **–** |
| **22** | **194** | **7.4** | **300** | **131** | **-120** | **10.4** | **15.4** | **1.67** | **51.5** | **1.62** | **1.27** | **1** | **ATD** | **–** | **–** |
| 23 | 194 | 7.0 | 250 | 81 | -180 | – | – | – | – | – | – | – | ATD | – | – |
| 24 | 204 | 5.4 | 304 | 74 | -130 | – | – | – | – | – | – | – | ATD | – | – |
| **25** | **204** | **4.9** | **233** | **63** | **-100** | **9.4** | **7.9** | **9.98** | **27.8** | **1.20** | **1.09** | **2** | **ATD** | **–** | **–** |
| **26** | **204** | **8.0** | **300** | **131** | **-130** | **9.7** | **13.1** | **7.72** | **49.0** | **1.27** | **1.12** | **2** | **ATD** | **–** | **–** |
| **27** | **204** | **8.1** | **300** | **131** | **-160** | **9.8** | **13.8** | **7.58** | **48.4** | **1.07** | **1.02** | **2** | **ATD** | **–** | **–** |
| 28 | 194 | 5.1 | 304 | 75 | -120 | 10.1 | 3.0 | 1.60 | 23.6 | 1.67 | – | 4 | – | – | – |
| 29 | 194 | 6.5 | 235 | 66 | -140 | 10.9 | 4.9 | 2.04 | 13.2 | 1.84 | – | 6 | SA | – | – |
| **30** | **194** | **7.6** | **300** | **131** | **-140** | **10.9** | **10.5** | **1.65** | **53.4** | **1.88** | **–** | **2** | **SA** | **–** | **–** |
| **31** | **194** | **7.5** | **300** | **131** | **-150** | **10.5** | **5.9** | **1.78** | **53.3** | **1.95** | **–** | **4** | **SA** | **–** | **–** |
| **32** | **194** | **7.6** | **300** | **131** | **-140** | **10.2** | **6.8** | **1.74** | **58.7** | **1.34** | **–** | **2** | **ATD-SA** | **–** | **–** |
| **33** | **194** | **7.6** | **300** | **139** | **-120** | **9.3** | **17.6** | **1.55** | **60.4** | **1.34** | **–** | **2** | **ATD-SA** | **–** | **–** |
| 34 | 189 | 7.3 | 306 | 136 | -140 | 10.6 | 13.6 | 0.78 | 21.1 | 1.84 | – | 4 | – | – | – |
| **35** | **189** | **7.3** | **305** | **137** | **-140** | **9.5** | **22.8** | **0.73** | **60.9** | **1.88** | **–** | **3** | **SOA** | **–** | **–** |
| **36** | **189** | **7.3** | **302** | **134** | **-150** | **–** | **–** | **0.58** | **0.26** | **1.95** | **–** | **–** | **SOA-H** | **–** | **–** |
| **37** | **189** | **7.2** | **301** | **132** | **-140** | **–** | **–** | **0.79** | **0.47** | **1.34** | **–** | **–** | **SOA-H** | **–** | **–** |
| **38** | **189** | **7.0** | **301** | **135** | **-120** | **–** | **–** | **0.72** | **0.32** | **1.34** | **–** | **–** | **SOA-H** | **–** | **–** |
| 39 | 224 | 6.1 | 300 | 71 | -120 | 13.2 | 0.6 | 112 | 1.3 | 1.24 | – | – | – | -0.01 | 0.4 |
| 40 | 224 | 7.6 | 234 | 65 | -120 | 13.3 | 3.3 | 129 | 14.8 | 1.18 | – | 2 | ATD | -0.02 | 0.7 |
| **41** | **224** | **8.9** | **300** | **134** | **-240** | **13.4** | **7.4** | **103** | **31.4** | **1.24** | **1.18** | **9** | **ATD** | **-0.03** | **1.7** |
| **42** | **224** | **8.4** | **300** | **130** | **-160** | **12.2** | **9.1** | **121** | **44.0** | **1.23** | **1.18** | **10** | **ATD** | **-0.04** | **2.7** |
| **43** | **224** | **8.5** | **300** | **130** | **-130** | **10.7** | **7.6** | **128** | **40.3** | **1.12** | **1.11** | **9** | **ATD** | **-0.04** | **2.0** |
| 44 | 204 | 8.0 | 300 | 71 | -300 | 12.8 | 3.8 | 7.96 | 30.8 | 1.71 | – | 4 | – | – | – |
| **45** | **205** | **8.3** | **300** | **134** | **-140** | **12.7** | **7.3** | **7.79** | **34.0** | **1.45** | **1.35** | **4** | **SA** | **–** | **–** |
| **46** | **204** | **5.5** | **301** | **74** | **-130** | **12.6** | **7.7** | **8.32** | **34.0** | **1.20** | **1.09** | **2** | **SA** | **–** | **–** |
| **47** | **204** | **5.2** | **233** | **64** | **-120** | **12.7** | **7.2** | **10.1** | **27.5** | **1.17** | **1.09** | **2** | **SA** | **–** | **–** |
| **48** | **204** | **7.6** | **301** | **132** | **-150** | **12.6** | **11.3** | **8.06** | **48.5** | **1.12** | **1.04** | **2** | **SA** | **–** | **–** |

Table S3: All adiabatic expansion experiments during IsoCloud 4; the 28 experiments used in this analysis are shown in boldface. Each subsection delineates experiments performed during a single day. Columns show: ID - experiment number; $T_0$(K) - initial temperature before pumps turn on; $\Delta T$(K) - change in temperature during an experiment; $p_0$ (hPa) - initial pressure; $\Delta p$ (hPa) - change in pressure during an experiment; $R_{vD,0}$ - initial isotopic ratio (in terms of natural abundance); $\delta R_{vD,0}$ - change in isotopic ratio due to an experiment; $r_{v,0}$ (ppmv) - initial mixing ratio of $H_2O$; $\delta r_{v,0}$ (ppmv) - change in mixing ratio due to ice deposition; $w_{eff}$ - cooling rate expressed as an effective updraft speed [23]; $S_{max}$ - maximum saturation during an experiment; $\bar{S}_i$ - deposition weighted average saturation; $d_{avg}$ ($\mu m$) - mean diameter; IN - ice nuclei; abbreviated as ATD (Arizona Test Dust), SA (sulfate aerosols), SOA (secondary organic aerosols), and SOA-H (secondary organic aerosols plus nitric acid); $R_{vD,dev}$ - maximum fractional correction on isotopic ratio due to backscattering during a pumpdown; $r_{dev}$ - maximum fractional correction on mixing ratio of $H_2O$ due to backscattering correction during a pumpdown. (The maximum corrections in last two columns generally exceed those in the experimental regions analyzed.)

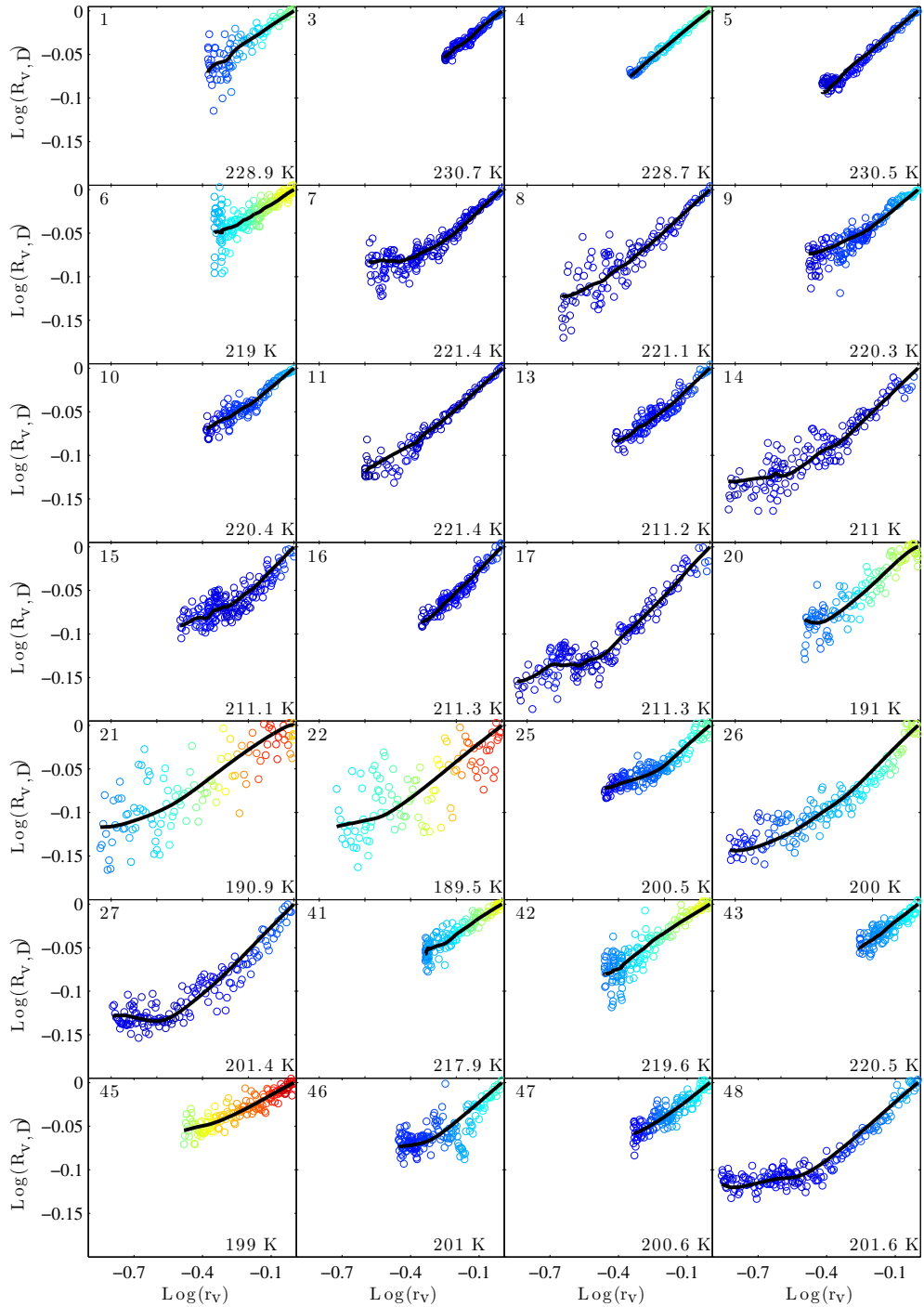

Figure S4: Isotopic evolution in all adiabatic expansion experiments used in this analysis. Points represent measured data during the defined analysis region, coloured by degree of supersaturation with respect to ice. (Colour scale from blue to red indicates $S_i$ of 0.95 to 1.5.) Black line shows the best-fit model of isotopic evolution, as described in S4. Time evolution in each experiment proceeds from top right to lower left, with the initial slope giving $\alpha_{\text{eff}} - 1$. Temperature at lower right in each panel is the initial temperature in each analysis region (several degrees lower than the pumpdown start temperature listed in Table S3).

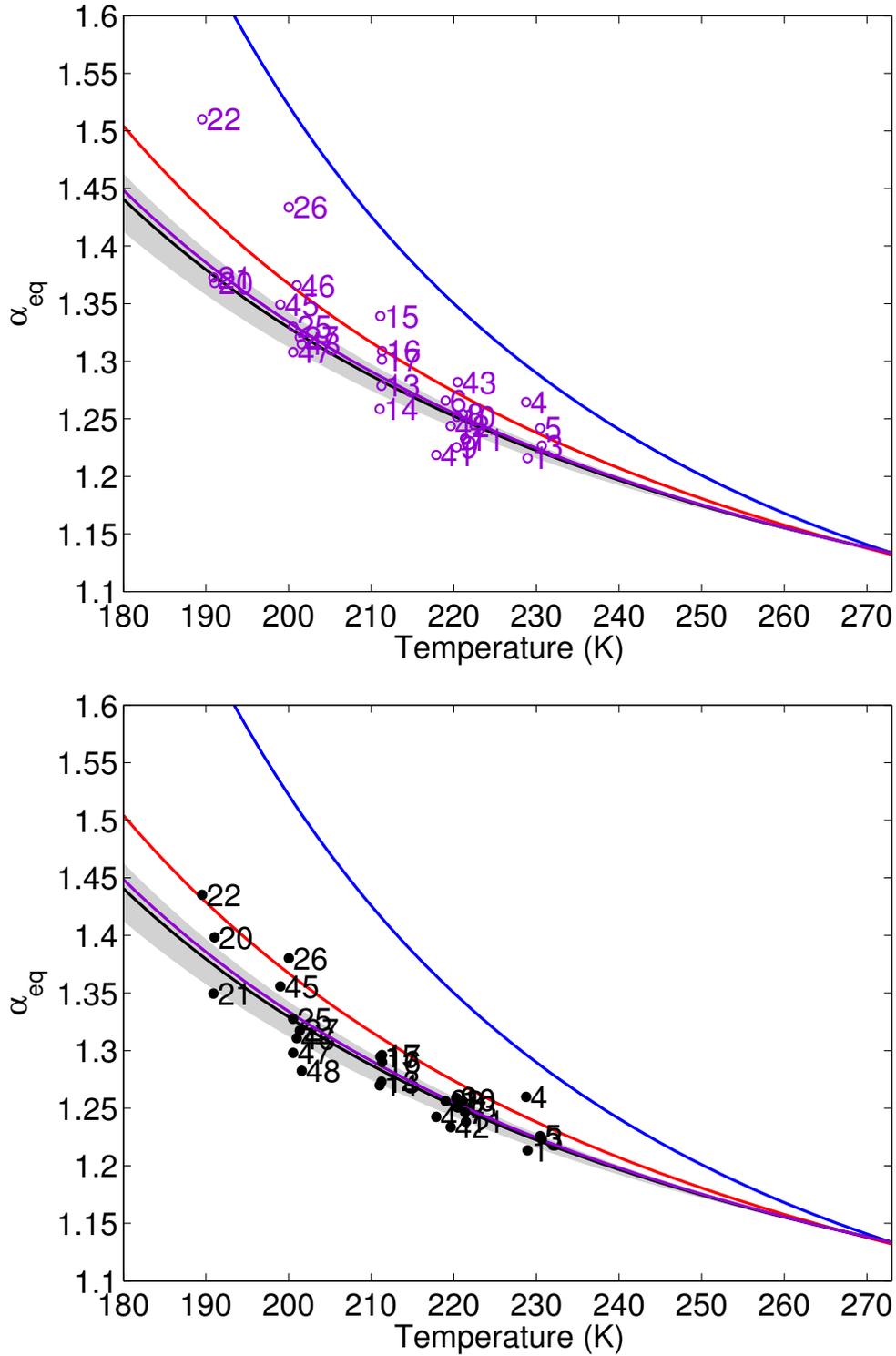

Figure S5: Fits for $\alpha_{\mathrm{eq}}$, identified by experiment number. **Top:** the 2-parameter case, and **bottom:** the 1-parameter case. Black and purples lines show the corresponding global fits, and blue and red lines show the E13 and M67 parameterizations, respectively. Fitting procedure is described in text and S4-S5. See Figure S1 for error bars. Experimental clustering is tighter in the 1-parameter case. Because the data contain insufficient information to determine two parameters independently, fit degeneracy in the 2-parameter case produces strong correlation between derived values for the equilibrium fractionation factor $\alpha_{\mathrm{eq}}$ and the isotopic composition of wall outgassing $R_w$.

## S3  Isotopic model for expansion experiments

We model the evolving isotopic composition in the vapour during a pseudo-adiabatic expansion experiment from the conservation of mass mixing ratio of the heavier isotopologue of the total water content inside AIDA, given a sink due to ice deposition and a source from wall outgassing during pumpdowns. Ice deposition is assumed to proceed with an unknown fractionation factor $\alpha_{\text{eff}}$ that is dependent on both temperature and saturation: $\alpha_{\text{eff}} = \alpha_k \alpha_{\text{eq}}$, where $\alpha_{\text{eq}}$ is the temperature-dependent equilibrium fractionation factor and $\alpha_k$ the kinetic modification. The isotopic composition of the wall contribution is an unknown $R_w$, fit either as a free parameter or as a function of $\alpha_{\text{eq}}$ (See S4.1.1).

We denote the mass mixing ratio of vapour and ice phases as $r_v$ and $r_i$ (and those of the heavy isotoplogue as $r'_v$ and $r'_i$; we use the prime symbol throughout to denote the isotopically substituted quantity). The isotopic ratios for vapour and ice are then $R_v = r'_v/r_v$ and $R_i = r'_i/r_i$, respectively.

Conservation of mass of the abundant isotopologue can be written simply as:

$$\frac{dr_v}{dt} = -P_{\text{vi}} + S_{\text{wv}}, \tag{1}$$

$$\frac{dr_i}{dt} = P_{\text{vi}}, \tag{2}$$

where $P_{\text{vi}}$ and $S_{\text{wv}}$ stand for the depositional production of ice from vapour and the source of vapour from wall outgassing, respectively. (Both are $> 0$ in normal conditions.) The mixing ratio of total water $r_{\text{tot}}$ (vapour plus ice) in the chamber then evolves as $dr_{\text{tot}}/dt = S_{\text{wv}}$. The wall contribution can be assumed to be directly into the vapour phase.

Conservation of the heavy isotopologue content gives:

$$\frac{dr'_v}{dt} = -\alpha_{\text{eff}} R_v P_{\text{vi}} + R_w S_{\text{wv}}, \tag{3}$$

where $R_v$ and $R_w$ are the isotopic ratios of chamber vapour (measured) and of vapour outgassing from the wall (unknown). The first term on the right-hand side of Eq. (3) describes the tendency of the vapour during depositional growth of ice. The isotopic ratio of the surface of an ice crystal growing by deposition is $R_i^{(s)} = \alpha_{\text{eff}} R_v$, where $\alpha_{\text{eff}}$ is the temperature- and saturation-dependent effective fractionation factor [24]. $\alpha_{\text{eff}}$ differs from $\alpha_{\text{eq}}$ when diffusion-limited growth produces kinetic fractionation between the vapour and the ice surface.

Expanding Eq. (3) using the definition of $R_v$ and substituting into Eq. (1) yields the model for the evolution of vapour isotopic ratio in the chamber during the pumpdown:

$$\boxed{\frac{dR_v}{dt} = -\left(\alpha_{\text{eff}} - 1\right) R_v \frac{P_{\text{vi}}}{r_v} + (\gamma - 1) R_v \frac{S_{\text{wv}}}{r_v}.} \tag{4}$$

where $\gamma \equiv R_w/R_v$. The first term on the RHS describes distillation due to ice depositing during the pumpdown, and the second term describes enrichment due to wall outgassing. $\gamma$ plays a role symmetric to $\alpha_{\text{eff}}$, i.e. it describes the enrichment of outgassing vapour relative to that of chamber vapour. In the limit that $R_w \to R_v$, Eq. (4) reduces to the equation for simple Rayleigh distillation.

In a few experiments, we find that $S_{\text{wv}} < 0$, i.e. total water mixing ratio decreases for short periods. Such cases would correspond to situations where the walls are not outgassing, but rather ice is depositing both on chamber walls and on crystals inside the chamber. In these instances we modify the model in Eq. (4) by reassigning variables as follows:

$$-P_{\text{vi}} \to -P_{\text{vi}} + S_{\text{wv}}, \tag{5}$$
$$S_{\text{wv}} \to 0. \tag{6}$$

In the experiments described here, isotopic doping does not affect retrieval of the equilibrium fractionation factor. Because doping levels are only $\sim$10-20 times natural abundance, the heavier isotopologue is still dilute with respect to the lighter isotopologue. In the dilute case, by Raoult's Law, the equilibrium fractionation factor will be equal to the ratio of HDO to $H_2O$ partial pressures over pure HDO and $H_2O$, respectively.

The kinetic modification of the fractionation factor can be significant in some experiments, particularly those at cold temperatures and high supersaturation with respect to ice. We model the kinetic fractionation factor as linked to the equilibrium factor at ice deposition by the relationship:

$$\alpha_k = \frac{S_i}{\alpha_{\text{eq}} \cdot g\,(S_i - 1) + 1} \tag{7}$$

where $g$ is a coefficient that controls the magnitude of kinetic modifications. In the standard diffusive model of [24], $g$ is equal to the ratio of molecular diffusivities of $H_2O$ and HDO ($g = \text{d} = D_v/D_v'$). In a variation of the model that includes surface kinetic effects, $g$ is a more complex term that is a function of not only d but ice crystal diameter, thermal velocities, and the ratio of deposition coefficients for $H_2O$ and HDO [25]. This model is discussed in detail in S6. In all analyses here, we omit effects related to the ratio of ventilation coefficients (negligible for small crystals and low temperatures, see [24, 26]), thermal impedance (effects on $\alpha_{\text{eq}}$ of order $10^{-3}$ or less), and corrections due to variation of thermodynamic quantities across the thermal boundary layer (small in all cases, see [27]).

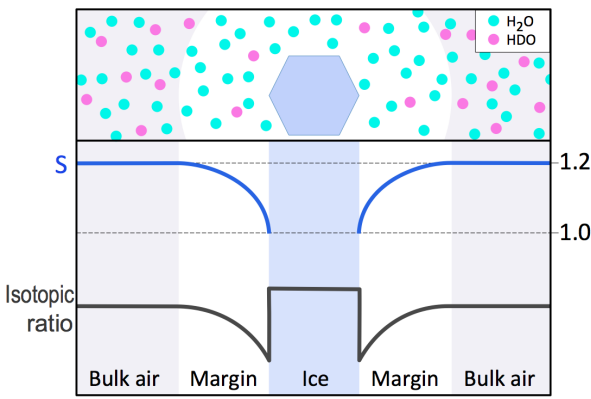

Figure S6: Cartoon demonstrating the effects of diffusion limitation on isotopic fractionation during ice growth. When diffusion becomes limiting, the effective fractionation becomes lower than in the equilibrium case because vapour isotopic composition is lower in the near field than in the bulk gas. This isotopic gradient arises primarily through two mechanisms: preferential uptake of the heavier isotopologue ($\alpha_{\text{eq}}$) and lower diffusivity of the heavier isotopologue (d). In more complex models, differences in deposition coefficients between isotopologues can also play a role. See S6 for further discussion.

# S4 Fitting protocol: individual experiments

## S4.1 Region selection

To determine the isotopic fractionation factor, it is important to select a region of the experiment where the deposition of vapour as ice is maximal and the wall contribution is minimal, since the isotopic composition of the wall contribution is an unknown that we are approximating as constant. Most ice growth occurs in the first few minutes of each experiment, while outgassing from the chamber walls becomes progressively more important as the experiment proceeds. Restricting the analyzed region to a segment near the beginning of each experiment therefore maximizes signal from the isotopic fractionation and minimizes contributions from the wall: the two terms on the right side of Eq. (4). For each experiment (defined as beginning at the onset of pumping) we select the analysis region using the criteria:

*Beginning*
- At least 2% of the initial vapour must have deposited as ice
- The isotopic ratio must have changed by at least 0.5% of its initial value

*End*
- The outgassing rate must be less than or equal to the ice deposition rate
- The cumulative flux from the chamber walls must be less than 15% of the initial vapour

Regions selected based on these criteria range from 54 to 223 seconds long, with a mean length of 136 seconds. Region start times begin between 30 and 200 seconds after the onset of pumping, so that the analysis region does not include the first several degrees of cooling. Start times are later at the coldest temperatures because of higher nucleation thresholds and slower rates of ice growth.

## S4.2 Model implementation

The fractionation factors are estimated for each pumpdown by minimizing the difference between observed and modeled vapour isotopic ratios, with the isotopic ratios modeled according to Eq. (4). To arrive at a closed form solution, we parametrize $\alpha_{\text{eq}}$ and $\gamma$ as follows:

- **equilibrium fractionation factor:** $\alpha_{\text{eq}} = \alpha_0 + \left.\frac{\partial \alpha_{\text{eq}}}{\partial T}\right|_0 (T - T_0)$, where $T_0$ is the temperature at the beginning of the pumpdown. This linearization is acceptable because the variation in $\alpha_{\text{eq}}$ over the course of each pumpdown is expected to be a small fraction of the dynamic range of $\alpha_{\text{eq}}$ over all pumpdowns. Temperature changes during pumpdowns are of order 5-9 K (2-5 K for regions analyzed), producing change in $\alpha_{\text{eq}}$ of less than $\sim 5\%$). This parameterization introduces a dependence of $\alpha_{\text{eq}}$ on the unknown $\frac{\partial \alpha_{\text{eq}}}{\partial T}$ and so mandates an iterative solution: we start from an initial guess of $\frac{\partial \alpha_{\text{eq}}}{\partial T}$, solve for $\alpha_0$ in individual experiments, fit for the global temperature dependence, update $\frac{\partial \alpha_{\text{eq}}}{\partial T}$ based on that solution, and iterate to convergence.

- **relative isotopic enrichment of outgassing vapour:** $\gamma = R_w/R_v$, with $R_v$ the measured vapour isotopic composition, and $R_w$ the isotopic composition of wall outgassing assumed constant during the analyzed period of each experiment. In reality the composition of the ice layers coating the chamber walls may show some structure, but the outgassing composition likely varies sufficiently slowly to justify its approximation as constant. We set the constant $R_w$ in two ways:

    - **2-parameter fit:** We fit $R_w$ as a free parameter independent from $\alpha_{\text{eq}}$. In practice, the pumpdowns do not contain sufficient information to fully constrain $R_w$ and $\alpha_{\text{eq}}$ separately, and the fitted values of the two variables are correlated. We use this fit as a reality check on results, and to estimate error bars on $\alpha_{\text{eq}}$ for individual experiments.
    - **1-parameter fit:** $R_w = \alpha_{\text{eq}} \cdot R_{v0}$. We assume the wall outgassing composition is that of an ice layer in equilibrium with chamber vapour before the pumpdown. $R_w$ is then tied to the equilibrium fractionation factor $\alpha_{\text{eq}}$ and the measured pre-pumpdown isotopic ratio in chamber vapour $R_{v0}$. Data points in manuscript Figure 2 and stated coefficients for $\alpha_{\text{eq}}$ are taken from this fit.

*Treatment of other model variables.* While we fit on the raw 1-second measurements of isotopic ratio, we use an effectively smoothed version of the measured evolving water vapour and total water to compute the terms $P_{\text{vi}}/r_v$ and $S_{\text{wv}}/r_v$ in Eq. (4). That is, we take the low-order trend by computing the lowest SSA eigenmodes of the relevant measured time series ([28], see S4.4 for further discussion of SSA method). The modeled evolution of the isotopic ratio of vapour in the chamber is then determined by integrating Eq. (4) using a Runge-Kutta method with adaptative stepsize control, and the modeled isotopic ratio is fit to the raw measured isotopic ratio as described in the next section.

### S4.3   Parameter and uncertainty estimation

We estimate parameters by the method of maximum likelihood. We denote the vector of parameters to estimate as $\boldsymbol{\theta}$. In the 2-parameter fit, $\boldsymbol{\theta} = [\theta_1, \theta_2]^T$ with $\theta_1 = \alpha_0$ and $\theta_2 = \gamma_0$. In the 1-parameter fit, $\boldsymbol{\theta} = [\theta_1]$ with $\theta_1 = \alpha_0$.

Measured and modeled isotopic ratios are related as follows:

$$\mathbf{R}^{\text{obs}} = \mathbf{R}^{\text{m}}(\boldsymbol{\theta}) + \boldsymbol{\varepsilon} \tag{8}$$

where $\mathbf{R}^{\text{obs}}$ are measured and $\mathbf{R}^{\text{m}}$ modeled isotopic ratios and $\boldsymbol{\varepsilon}$ are inferred measurement errors, discussed in S4.4. Because the errors are reasonably independent and normally distributed, the likelihood of $\boldsymbol{\theta}$ (that is, the probability of observing $\mathbf{R}^{\text{obs}}$ given $\boldsymbol{\theta}$) can be written as:

$$\begin{aligned}
L(\boldsymbol{\theta}) &= p\left(\mathbf{R}^{\text{obs}}|\boldsymbol{\theta}\right) \\
&= \frac{1}{(2\pi)^{\frac{N}{2}}\sqrt{\det(\mathbf{C}_{\boldsymbol{\varepsilon}})}} \\
&\quad \times \exp\left[-\frac{1}{2}\left(\mathbf{R}^{\text{obs}} - \mathbf{R}^{\text{m}}(\boldsymbol{\theta})\right)^T \mathbf{C}_{\boldsymbol{\varepsilon}}^{-1}\left(\mathbf{R}^{\text{obs}} - \mathbf{R}^{\text{m}}(\boldsymbol{\theta})\right)\right]
\end{aligned} \tag{9}$$

where $N$ is the dimension of $\mathbf{R}^{\text{obs}}$ (or $\mathbf{R}^{\text{m}}$) and $\mathbf{C}_{\boldsymbol{\varepsilon}} = \mathrm{E}\left[\boldsymbol{\varepsilon}\boldsymbol{\varepsilon}^T\right] = \mathrm{diag}\left(\sigma_i^2\right)$ is the covariance matrix of measurement errors. (See S4.4 for discussion of estimating measurement error $\sigma_i$.) The maximum likelihood estimator of $\boldsymbol{\theta}$ is then the value $\hat{\boldsymbol{\theta}}$ that maximizes $L(\boldsymbol{\theta})$. (We use the "hat" symbol throughout to denote an estimated quantity.) In practice, we minimize $-\mathcal{L}(\boldsymbol{\theta}) = -\log L(\boldsymbol{\theta})$, which is equivalent to minimizing the following quantity:

$$\frac{1}{2}\sum_{i=1}^{N}\left(\frac{R_i^{\text{obs}} - R_i^{\text{m}}(\boldsymbol{\theta})}{\sigma_i}\right)^2 \tag{10}$$

The minimization of $-\mathcal{L}(\boldsymbol{\theta})$ is carried out using a quasi-Newton algorithm.

Uncertainty in retrieved parameters $\boldsymbol{\theta}$ is estimated using Fisher matrix theory. The likelihood $L(\boldsymbol{\theta})$ is asymptotically Gaussian near its maximum, i.e. it can be written:

$$L(\boldsymbol{\theta}) \propto \exp\left[-\frac{1}{2}\left(\boldsymbol{\theta}-\hat{\boldsymbol{\theta}}\right)^T \mathbf{C}_{\boldsymbol{\theta}}^{-1}\left(\boldsymbol{\theta}-\hat{\boldsymbol{\theta}}\right)\right] \tag{11}$$

where $\mathbf{C}_{\boldsymbol{\theta}}$ is the covariance matrix of the estimated parameters. (The inverse of $\mathbf{C}_{\boldsymbol{\theta}}$ is called the Fisher information matrix.) $\mathbf{C}_{\boldsymbol{\theta}}$ is computed as the inverse of the Hessian of the log-likelihood:

$$\mathbf{C}_{\boldsymbol{\theta}} = -\left(\nabla_{\boldsymbol{\theta}\boldsymbol{\theta}}\mathcal{L}\right)^{-1} = -\left(\nabla_{\boldsymbol{\theta}\boldsymbol{\theta}}\log L\right)^{-1} \tag{12}$$

(The Hessian matrix is computed using centered finite differences.) The diagonal of $\mathbf{C}_{\boldsymbol{\theta}}$ then contains the variance estimates of the parameters $\theta_i$. That is, $\mathbf{C}_{\boldsymbol{\theta}}$ produces an estimate of the uncertainty $\sigma_n^{\alpha}$ in each retrieval of an equilibrium fractionation factor $\hat{\alpha}_n$. These uncertainties are shown as the $2\sigma$ error bars in Figure S1 and manuscript Figure 3.

Comparison of the estimated $\sigma_n^{\alpha}$ for individual experiments and the distribution of derived fractionation factors $\hat{\alpha}_n$ around the fit to all experiments suggests that these error estimates are appropriate for the low-temperature experiments, in which signal-to-noise is the dominant factor driving uncertainty, but are underestimates for the warmer experiments (T> 210 K), in which un-modeled systematic errors (likely driven by chamber inhomogeneities) dominate. We do not attempt to estimate additional contributions to experimental error, but note that true uncertainties may be a factor of three larger at the warmer end of the experimental temperature range.

### S4.4 Measurement error estimation

The likelihood function used to estimate parameters for each pumpdown (Eq. 10) requires an estimation of the intrinsic uncertainty $\sigma_i$ in the isotopic ratio measurements. Determining this uncertainty would be trivial if its only source were measurement precision and instrument characteristics were fixed; in this case measurement errors could simply be estimated from instrument performance in static conditions. Error estimation during pumpdowns is more challenging because instrument performance can be altered by vibrations and attenuation of signal, and because chamber inhomogeneities increase during pumpdowns. Effective measurement errors are therefore not constant during experiments. Since we are interested in changes in isotopic ratio rather than absolute changes in concentrations, uncertainties in spectroscopic parameters and most other factors affecting accuracy do not affect results.

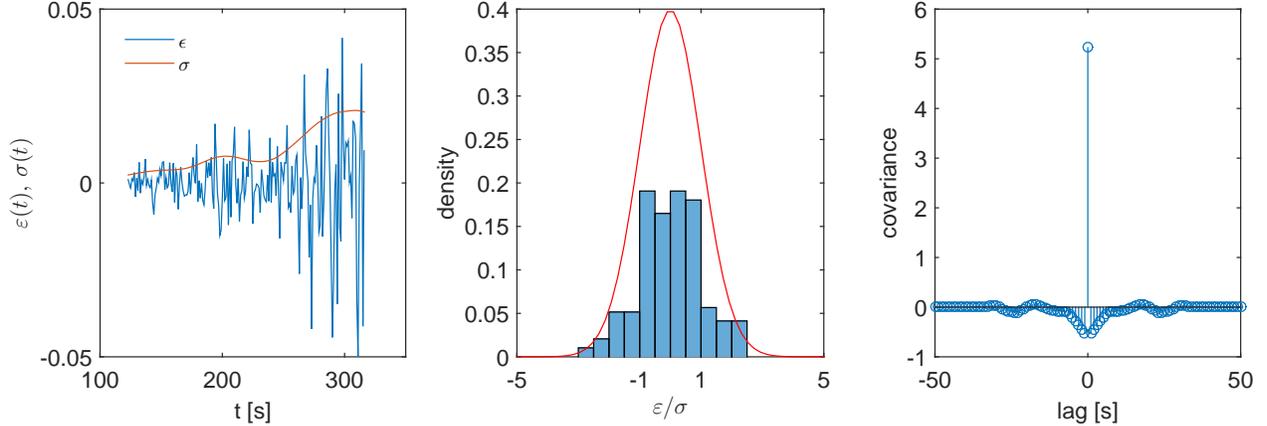

Figure S7: Example of estimated noise on isotopic ratio measurements for experiment #6. **Left:** The reconstructed noise from the trailing SSA eigenmodes and their associated standard deviation. Note that noise grows over the course of the pumpdown. **Center:** The standardized error is reasonably normally distributed. **Right:** The reconstructed noise is reasonably decorrellated in time.

To estimate effective measurement error in a timeseries of isotopic ratio measurements, we assume that the effective error is manifested as deviations from a smoothly evolving isotopic ratio. While the fit for $\hat{\alpha}_n$ is performed on raw isotopic measurements, we estimate errors by taking a Singular Spectrum Analysis (SSA) [28] on that raw timeseries. SSA decomposes the timeseries into a sum of eigenmodes (temporal principal components), each of them accounting in decreasing order for a fraction of the variance of the original time series. We use the first few leading eigenmodes to construct the assumed smoothly evolving timeseries, and consider the trailing eigenmodes to represent the noise $\varepsilon_i$ on the measured isotopic ratio. The rank at which we set this eigenmode separation is determined by the test that errors must be reasonably time-decorrelated.

Note that this method cannot distinguish one important experimental artifact: low-frequency fluctuations due to chamber inhomogeneities, which are correlated in time. Isotopic measurements probe only the volume intersected by our infrared laser beam; chamber air circulating across the beam path results in small but important real fluctuations in temperature, water vapour content, and ice crystal number density, with timescale $\sim 30$ s.

Because effective noise characteristics change during a pumpdown (Fig. S7, left), we cannot simply take measurement uncertainty as the sample standard deviation of the noise $\varepsilon_i$, but instead must construct a standard error $\sigma_i$ that changes over the $i$ datapoints of the pumpdown. For this estimation, we use the fact that the noise covariance matrix $\mathrm{E}\left[\boldsymbol{\varepsilon}\boldsymbol{\varepsilon}^T\right] = \mathrm{diag}\left(\sigma_i^2\right)$, so that $\sigma_i^2$ can be taken as the leading SSA eigenmode of the $\varepsilon_i^2$ time series. We then test the validity of our approach for estimating $\varepsilon_i$ and associated standard deviation $\sigma_i$ by checking the distribution of the standardized errors $\varepsilon_i/\sigma_i$ in each pumpdown. Because our approach models the likelihood of our parameters as a multivariate normal distribution of variance $\sigma_i$, values of $\varepsilon_i/\sigma_i$ during a pumpdown should be normally distributed with variance 1 and decorrelated in time. We find that they are indeed sufficiently well behaved with respect to our basic assumptions (Fig. S7). In the next section, we find that the $\sigma_i$ are also optimal, in the sense that rescaling only marginally improves likelihoods.

### S4.4.1 Error rescaling

As a check on our error estimation procedure, we also test the possibility that the covariance matrix of measurement errors $\mathbf{C}_\varepsilon$ previously derived is known only up to a scaling factor. That is, we consider whether the standard deviations of measurement errors may in fact be $s\sigma_i$, where $s$ is a global scaling factor attached to each pumpdown. In that case, $\mathbf{C}_\varepsilon$ would be known up to $s^2$. In the previous section, we assumed $s = 1$. We now determine $s$ by treating it as one of the parameters estimated by maximizing the likelihood function. For instance, with $\theta_1 = \alpha_0$ and $\theta_2 = x_0$, we set $\theta_3 = s$. The minimization of $-\mathcal{L}(\boldsymbol{\theta}) = -\log L(\boldsymbol{\theta})$ is now equivalent to minimizing the quantity:

$$\sum_{i=1}^{N} \log(\theta_3 \sigma_i) + \frac{1}{2} \sum_{i=1}^{N} \left( \frac{R_i^{\text{obs}} - R_i^{\text{m}}(\theta_1, \theta_2)}{\theta_3 \sigma_i} \right)^2. \tag{13}$$

As can be seen in Fig. S8, rescaling the errors improves the likelihood only marginally, meaning this operation is essentially superfluous for estimating the fractionation factors and their associated standard errors. These results imply that the characterization of errors described in Sec. S4.4 is already nearly optimal and cannot be much improved.

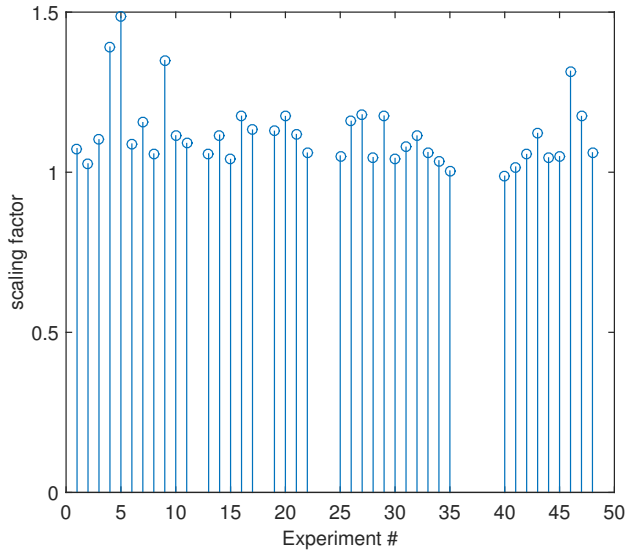

Figure S8: The derived scaling factor s for error for all experiments fits to close to 1, demonstrating that the noise reconstruction from the SSA eigenmodes provides a good estimation of the covariance matrix of measurement errors (likelihood is only marginally improved by rescaling). This means that the uncertainty on $\alpha_0$ for individual pumpdowns is sufficiently well-estimated using the minimization of the likelihood function without the inclusion of an additional scaling factor.

## S5 Fitting protocol: global fit for temperature dependence

### S5.1 Global fit procedure

After repeating the estimation procedure described above (S4.3) for all $n$ experiments to obtain individual estimates for the fractionation factors $\hat{\alpha}_n$ at temperatures $T_n$, and corre-

sponding uncertainties $\sigma_n^\alpha$, we then fit for the temperature dependence of $\alpha_{\text{eq}}$, assuming that it follows the form:

$$\log \alpha_{\text{eq}} = a_0 + \frac{a_1}{T^2} \quad (14)$$

A $1/T^2$ temperature dependence was also assumed in the M67 parameterization [4].

Eq. 14 stems from quantum mechanical considerations [2, 29], and is derived by expanding the partition functions for isotopically substituted and non-substituted species under the harmonic approximation for the force field potentials. A more general form including the effects of anharmonicity would be:

$$\log \alpha_{\text{eq}} = \delta_0 + \frac{\delta_1}{T} + \frac{\delta_2}{T^2} \quad (15)$$

In most isotopic systems, the anharmonic term $\delta_1/T$ can be neglected at high temperatures [30], but in the case of hydrogen partitioning the effects of anharmonicity and non-classical rotation are important. It is not straightforward to determine the appropriate functional form for the isotopic fractionation factor for HDO to $H_2O$, but our measurements suggest that Eq. (14) sufficiently represents the functional relationship, since the experimentally derived values for the fractionation factor for individual pumpdowns appear roughly linear in $\log(\alpha_{\text{eq}})$ vs. $1/T^2$ space.

To provide consistency with previously measured values for the fractionation factor at higher temperatures, which agree well, we further impose the constraint that we take the highest temperature measurement from [4] to be a true value, fixing $\alpha_0 = 1.143$ at $T_0 = -6$ °C. This constraint does not significantly affect the parameterization of $\alpha_{\text{eq}}$ over our experimental temperature range.

To estimate the temperature dependence, we make Eq. (14) linear with a change of variables to $x = \log \alpha_{\text{eq}}$ and $t = 1/T^2$: $x = a_0 + a_1 t$. Including the constraint at $T_0$ then yields:

$$x - x_0 = a_1 (t - t_0) \quad (16)$$

and estimating the temperature dependence $\delta_1$ becomes a linear estimation problem:

$$\mathbf{x} - \mathbf{x_0} = \mathbf{H} \cdot a_1 + \mathbf{w} \quad (17)$$

where $\mathbf{x} - \mathbf{x_0}$ describes fractionation values derived for individual experiments: $\mathbf{x} - \mathbf{x_0} = [0, x_1 - x_0, ..., x_N - x_0]^T$, and $\mathbf{H}$ the experimental temperatures: $\mathbf{H} = [0, t_1 - t_0, ..., t_N - t_0]^T$. $\mathbf{w}$ is the vector of errors on the determination of $\log \alpha_{\text{eq}}$; the covariance matrix of $\mathbf{w}$ is $\mathbf{C_w} = \text{diag}\left[(\sigma_n^\alpha/\hat{\alpha}_n)^2\right]$. (See S4.3 for discussion of $\hat{\alpha}_n$ and $\sigma_n^\alpha$.) Following standard theory, the best linear estimator of $a_1$ is:

$$\hat{a}_1 = \left(\mathbf{H}^T \mathbf{C_w}^{-1} \mathbf{H}\right)^{-1} \mathbf{H}^T \mathbf{C_w}^{-1} \cdot (\mathbf{x} - \mathbf{x_0}) \quad (18)$$

and its uncertainty is $(\sigma^{a_1})^2 = \left(\mathbf{H}^T \mathbf{C_w}^{-1} \mathbf{H}\right)^{-1}$, which can be used to build confidence intervals.

Note that during the estimation procedure, $\mathbf{C_w}$ is internally resized in order to better match the dispersion of the residuals. That is, $\mathbf{C_w}$ is rescaled by a factor such that the variance of the standardized residuals becomes 1 (compensating for the underestimation of uncertainties for individual experiments noted earlier). This rescaling does not affect the

estimated $\hat{a}_1$ but increases its estimated uncertainty $(\sigma^{a_1})^2$ by approximately a factor of three. The confidence intervals on $\alpha_{\text{eq}}$ shown throughout this work reflect this conservative estimate. The uniform rescaling is inappropriate given that uncertainties appear underestimated only for the warmer experiments, but sensitivity tests show that the resulting parametrization for $\alpha_{\text{eq}}$ is negligibly affected by the weighting of individual experiments. (See S7.3.)

## S5.2 Results for different treatment of $R_w$

Different assumptions about the isotopic composition $R_w$ of the wall outgassing flux produce different estimates of equilibrium fractionation $\hat{\alpha}_n$ for individual experiments, which in turn can lead to slightly different inferred temperature dependence for $\alpha_{\text{eq}}$. (See Fig. S5.) As described in S4.2, we use two different methods of fitting, each with desirable and undesirable features.

- In the 1-parameter case, we treat the wall as ice that has equilibrated with initial chamber vapour, ie. $\gamma = \alpha_{\text{eq}} R_{v0}^{obs}$. The only retrieved parameter is then the fractionation factor ($\alpha_0$). In this case, the estimated uncertainties on $\alpha_0$ for all experiments are implausibly optimistic, since the error bars reflect only measurement errors.

- In the 2-parameter case, we treat the isotopic composition of the wall outgassing flux $R_w$ (and so $\gamma = R_w/R_v^{obs}$) as an unknown that must be fit. Error bars in this case are considerably larger, reflecting additional uncertainty in $\gamma$. (The off-diagonal terms of the Fisher matrix are non-zero, indicating that the parameter determination errors for $\gamma$ and $\alpha_0$ are correlated.) The resulting uncertainty estimates $\sigma_n^\alpha$ arguably more closely approximate the true uncertainties on a determination of the fractionation factor. However, the dependence of the two parameters also produces fit degeneracy, with in many cases unphysically high values for $R_w$ and correspondingly large $\alpha_0$.

In this analysis we choose the 1-parameter fit as the default case, but make the further assumption that the uncertainty derived from the 2-parameter fit more closely approximates actual measurement uncertainty. Each choice produces a slightly different temperature dependence for $\alpha_{\text{eq}}$, given in Table S4 below.

| # | Assumption | $a_0$ | $a_1$ |
|---|---|---|---|
| (A) | 2 parameter ($\theta_1 = \alpha_0$, $\theta_2 = \gamma_0$) | -0.0619 | 13959 |
| (B) | 1 parameter ($\theta_1 = \alpha_0$, $\sigma_n^\alpha$ from (A)) | -0.0559 | 13525 |
| (C) | 1 parameter ($\theta_1 = \alpha_0$) | -0.0536 | 13364 |
| (M67) | Merlivat and Nief, 1967 | -0.0945 | 16289 |

Table S4: Parameters $a_0$ and $a_1$ obtained upon fitting $\log \alpha = a_0 + a_1/T^2$ to fractionation factors derived under various model assumptions, with M67 shown for comparison. Case (B) is that presented in the manuscript and shown in black in figures; case (A) is shown in purple.

## S6 Evaluation of kinetic models

Because the IsoCloud 4 experiments include a variety of mean supersaturations $\bar{S}_i$, conditions in which kinetic isotope effects should have differing influence, they allow testing the predictions of different models for the kinetic modification to fractionation. In all cases, we assume that kinetic isotopic effects can be represented by Eq. (7), i.e.:

$$\alpha_k = \frac{S_i}{\alpha_{\text{eq}} \cdot g\,(S_i - 1) + 1}$$

We test the effect of different proposed values for the parameter $g$, which is a function of d, the ratio of diffusivities of $H_2O$ and HDO, and, in some treatments, of surface effects including the ratio of deposition coefficients between isotopologues.

### S6.1 Kinetic fractionation models

We test two different treatments of kinetic isotopic effects.

**Diffusive flux model:** The default representation of kinetic modifications to fractionation in this work is the diffusive model of Jouzel and Merlivat [24], i.e. Eq. (7) with

$$g = \text{d}$$

**Diffusive plus surface-kinetics model:** Nelson [25] extends the diffusive flux model to include surface kinetic effects by setting

$$g = \text{d} \cdot \frac{1 + z'}{1 + z}$$

where $z = Z_S/Z_V$ is the ratio of the surface impedance to the vapour impedance, and as before, we denote quantities associated with the heavier isotopologue with a prime. If we assume spherical geometry for ice crystals, this expression can be rewritten as

$$g = \frac{\text{d}k + \text{yx}}{1 + k}$$

where x is the ratio of deposition coefficients for $H_2O$ and HDO (x=$\beta/\beta'$) and y is their ratio of thermal velocities ($\sqrt{19/18}$). The coefficient $k \equiv rv\beta/(4D_v)$, where $r$ is the ice crystal radius and $v$, $\beta$, and $D_v$ are the thermal velocity, deposition coefficient, and diffusion coefficient in air, respectively, for the abundant isotopologue $H_2O$. We make the simplifying assumption that $\beta = 1$; the value is unknown but is likely of order 1 [31]. We take the temperature- and pressure-dependent $D_v$ from Hall and Pruppacher 1976 [32]. In the limit $k >> 1$, the expression reduces to that of the diffusive model. In the experiments analyzed here, values for $k$ range from 2-15 across pumpdowns, essentially following the mean ice crystal diameter (2-14 $\mu$m). The greatest sensitivity to surface effects therefore occurs at the lowest temperatures, where ice crystal size and $k$ are smallest. The ratio of deposition coefficients x is essentially unknown; Nelson [25] suggests that its value could plausibly lie between 0.8 and 1.2.

### S6.2 Estimates of the isotopic diffusivity ratio

The value of d has been determined in a limited number of experiments, whose results are inconsistent within stated error bars. (Published values differ by as much as 3% while stated error bars are < 0.1%.) We summarize published results below in Table S5. In this work we take as our default for d the value from Cappa et al. 2003 [33].

Kinetic isotope effects in our experiments are predominantly driven by equilibrium fractionation during ice deposition, with differences in diffusivity playing only a minor role. In the diffusive flux model for kinetic isotopic effects given above, d plays an identical role to the equilibrium fractionation factor $\alpha_{eq}$: both appear only in the coefficient $\alpha_{eq}d$, and while d is only a few percent above 1, $\alpha_{eq}$ is ∼1.2-1.4 over the experimental temperature range. Nevertheless, *uncertainty* in d can still contribute significantly to uncertainty in kinetic effects.

| Measurement | T (°C) | T (K) | $D_{HDO}/D_{H_2O}$ | $D_{H_2O}/D_{HDO}$ (d) |
|---|---|---|---|---|
| Ehhalt and Knott, 1965 [34] | 20.0 | 293.2 | 0.9852 ± 0.003 | 1.0150 |
| Cappa et al., 2003 [33] | 20.0 | 293.2 | 0.9839 | 1.0164 |
| Merlivat, 1978 [35] | 21.0 | 294.2 | 0.9755 ± 0.0009 | 1.0251 |
| Luz et al., 2009 [36] | *-83.2* | *190.0* | *0.9573* | *1.0446* |
| – | 10.0 | 283.2 | 0.9720 ± 0.0005 | 1.0288 |
| – | 20.1 | 293.3 | 0.9775 ± 0.0005 | 1.0230 |
| – | 39.8 | 313.0 | 0.9798 ± 0.0005 | 1.0206 |
| – | 69.5 | 342.7 | 0.9841 ± 0.0003 | 1.0162 |

Table S5: Published estimates of the ratio of diffusivities for HDO and $H_2O$ in air. Values from Cappa et al. are derived from kinetic theory and given without uncertainties. Values for Luz et al. in regular font are measurements; in italics is the value that would be extrapolated at 190 K by an unweighted linear fit to these values: d=1.0807-1.901·$10^{-4}$·T, with T in Kelvin. (Only Luz et al. suggest a temperature dependence).

### S6.3 Tests of kinetic models

We test the validity of models of the kinetic isotope effect by examining whether, after correction for kinetic contributions to fractionation, the values for equilibrium fractionation retrieved from individual experiments ($\hat{\alpha}_n$) show dependence on supersaturation $S_i$. A dependence on saturation would be interpreted as the signature of under- or over-correction for kinetic effects. Absence of a trend would suggest that the magnitude of kinetic effects had been correctly estimated.

Because the experiments analyzed here are conducted at different temperatures, and equilibrium fractionation is a function of temperature, we do not consider the derived equilibrium fractionation factors $\hat{\alpha}_n$ directly. Instead, we consider their implied coefficients $\hat{a}_1^n$. Because the model for $\alpha_{eq}$ is constrained to include a measured value at $T = -6°C$, our global fit for the temperature-dependent $\alpha_{eq}$ can be fully described by this single coefficient $a_1$. Similarly, each $\hat{\alpha}_n$ for an individual pumpdown defines a single $\hat{a}_1^n$ that quantifies the temperature dependence of the equilibrium fractionation factor that would be implied by

that experiment alone. In M67, $a_1$ is 16289. In our global fit, $a_1$ is slightly lower, 13525. If all measurements were perfect, and the equilibrium fractionation factor exactly followed the $1/T^2$ dependence of Eq. (14), then each experiment would produce $\hat{a}_1{}^n$ exactly equal to this $a_1$. In reality, values of $\hat{a}_1^n$ derived from individual experiments scatter around this value. (See cartoon in Fig. S9.)

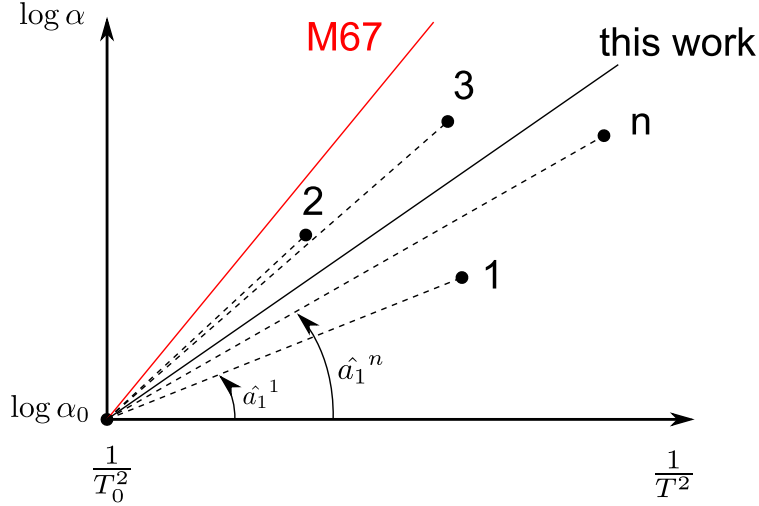

Figure S9: Cartoon to illustrate the slopes $\hat{a}_1{}^n$ that appear in manuscript Figure 3. $\hat{a}_1{}^n$ is the slope in $\log \alpha$ vs. $1/T^2$ space of the line connecting the assumed constraint point $(1/T_0^2, \log(\alpha_0))$ with an experimentally derived point $(1/T^2, \log(\hat{\alpha}_n))$. Individual experiments in this work generally produce $\hat{a}_1{}^n$s (slopes of dashed lines) below that of M67 (solid red) and therefore produce a weaker temperature dependence for the globally fit $\alpha_{\text{eq}}$ (solid black).

Any over- or under-correction for kinetic effects would produce a systematic trend in the $\hat{a}_1^n$s with supersaturation $S_i$. If $g$ in the kinetic model is incorrectly specified ($g \neq g_0$, where $g_0$ is the "true" value that would yield a perfect retrieval), then the resulting estimation of equilibrium fractionation $\alpha_{\text{eq}}$ would be biased, and that bias would depend on supersaturation:

$$\frac{\partial \alpha_{\text{eq}}}{\partial S_i} = (g - g_0)\, \alpha_{\text{eq}}^2 \tag{19}$$

This expression can be rewritten in terms of $a_1$ since

$$\ln \frac{\alpha_{\text{eq}}}{\alpha_0} = a_1 \left( \frac{1}{T^2} - \frac{1}{T_0^2} \right) \tag{20}$$

Combining equations 19 and 20 shows that $\partial \ln a_1 / \partial S_i$ is directly linked to the choice of $g$:

$$\frac{\partial \ln a_1}{\partial S_i} = (g - g_0) \frac{\alpha_{\text{eq}}}{\ln \left( \frac{\alpha_{\text{eq}}}{\alpha_0} \right)}, \tag{21}$$

The absence of trend with $S_i$ would therefore suggest that kinetic effects have been modeled correctly, i.e. $g = g_0$.

We conduct two tests to determine the validity of kinetic models and establish constraints on the parameters d and x. While we cannot constrain d and x simultaneously, we can examine their effects separately. First, we test a pure diffusive flux model ($g = d$) with different values of d. We fit for the optimal value of the diffusivity ratio d that eliminates supersaturation dependence, and determine bounds on potential values for d. We then test the diffusive-plus-surface-kinetic model with d set at the default value of Cappa et al. [33], and determine an optimum value and bounds for x given that assumption. That is, we assume that the value of the diffusivity ratio is known, and that any trend with saturation results from mis-estimation of surface kinetic effects due to an incorrect choice of the deposition coefficient ratio x. In both cases we fit for the trend of $\hat{a}_1{}^n$s vs. the deposition-weighted supersaturation

$$\bar{S}_i = \int P_{vi} S_i / r_v dt / \int P_{vi} / r_v dt \tag{22}$$

assuming a linear relationship: $a_1(S_i) = m \cdot S_i + b$. We seek the value of $g$ that yields zero slope ($m = 0$) by using a root-finding algorithm, and estimate bounds as described below. As in the fit for $\alpha_{\text{eq}}$, we weight individual experiments by their uncertainty $\sigma_n^\alpha$ (See S4).

The results of these tests yield bounds on diffusivity and deposition coefficient ratios d and x that are consistent with literature estimates of their plausible ranges. Figure S10 (analogous to manuscript Figure 3) shows results of the test on d with the pure diffusive model. The implied optimal value of d=1.01 is slightly lower than any measurement, but with bounds of $\sim \pm 4\%$ spanning the published values in Table S5: $0.97 < \text{d} < 1.05$. Figure S11 shows results of the test on x with the surface-kinetic model. Because the kinetic isotope effect is less sensitive to uncertainty in x, the test provides a looser constraint, with bounds on x exceeding $\pm 20\%$: $0.74 < \text{x} < 1.17$, spanning the range of 0.8-1.2 suggested by Nelson [25]. The central value of x=0.96 would imply that HDO molecules were slightly more likely to be accommodated into the crystal lattice than $H_2O$, but results are also consistent with x=1.

The looseness of these constraints stems from the small sample size of those IsoCloud 4 experiments sensitive to kinetic isotope effects. Only six experiments have deposition-weighted supersaturations above 1.2, three of those at temperatures low enough that signal-to-noise is poor and uncertainty in isotopic ratio measurements high. The small sample size also means that individual outliers can strongly bias results. In this analysis, we omit experiments #4, 26, and 48, three outlier experiments whose small nominal error bars leave them inconsistent with the global derived $\alpha_{\text{eq}}$ to greater than 5-$\sigma$ with respect to their estimated uncertainties; including these experiments would substantially alter the derived optimal values for d and x. Even assuming the uncertainties are underestimated by 3x, all three experiments remain outliers, with 2.5-$\sigma$ residual with respect to their estimated errors. Each of these experiments are also outliers with respect to the distribution in their respective temperature groups. The small sample size means that for d and x, IsoCloud results should be considered suggestive rather than conclusive. These results do however confirm the validity of kinetic models, and serve as a proof of concept of the approach. Future experimental campaigns targeted at kinetic isotope effects should be able to provide stronger constraints.

To estimate bounds on kinetic isotope effect parameters, we evaluate uncertainty $\sigma_g$ on estimating $g_0$ by propagating the uncertainty $\sigma_{\text{slp}}$ on the slope $\partial \ln a_1 / \partial S_i$. Because

experiments at different temperatures have different $\alpha_{\text{eq}}$, we know that uncertainty only to within some range. From Eq. (21), $\sigma_g$ must lie in the range:

$$\sigma_{\text{slp}} \cdot \min\{\frac{\ln\left(\frac{\alpha_{\text{eq}}}{\alpha_0}\right)}{\alpha_{\text{eq}}}\} \leq \sigma_g \leq \sigma_{\text{slp}} \cdot \max\{\frac{\ln\left(\frac{\alpha_{\text{eq}}}{\alpha_0}\right)}{\alpha_{\text{eq}}}\} \tag{23}$$

where the min and max values are over all experiments in the analysis. When assuming the diffusive model with $g =$ d to obtain constraints on d, we conservatively take the upper limit as the estimated uncertainty in d. When testing the surface-effect model with $g = (\text{d}k + \text{yx})/(1 + k)$, to obtain constraints on x, we can rewrite Eq. (21) as:

$$\frac{\partial \ln a_1}{\partial S_i} = (\text{x} - \text{x}_0) \frac{y}{1+k} \frac{\alpha_{\text{eq}}}{\ln\left(\frac{\alpha_{\text{eq}}}{\alpha_0}\right)} \tag{24}$$

where $\text{x}_0$ is the value of x that would yield the "true" retrieval of $\alpha_{\text{eq}}$. The uncertainty $\sigma_x$ on $\text{x}_0$ then satisfies the relationship:

$$\sigma_{\text{slp}} \cdot \min\{\frac{1+k}{y} \frac{\ln\left(\frac{\alpha_{\text{eq}}}{\alpha_0}\right)}{\alpha_{\text{eq}}}\} \leq \sigma_x \leq \sigma_{\text{slp}} \cdot \max\{\frac{1+k}{y} \frac{\ln\left(\frac{\alpha_{\text{eq}}}{\alpha_0}\right)}{\alpha_{\text{eq}}}\}, \tag{25}$$

We again conservatively take the upper limit as our estimate of uncertainty in $\text{x}_0$.

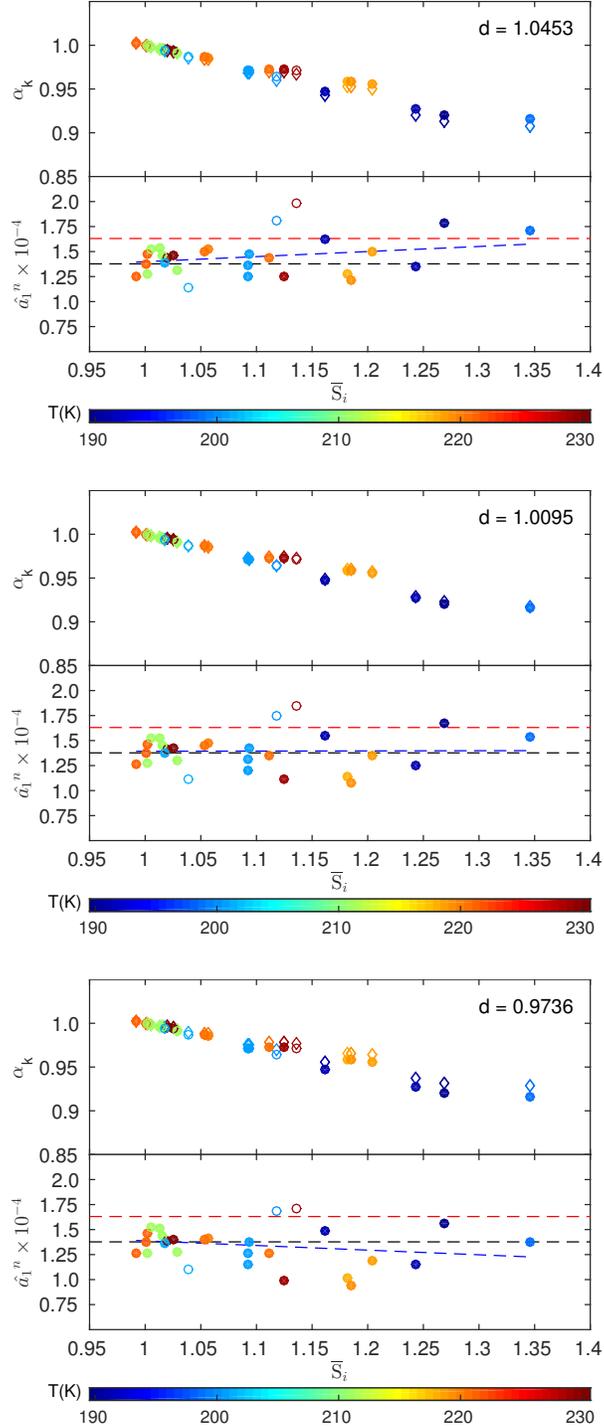

Figure S10: Effect of choice of d (ratio of isotopic diffusivities) on calculated kinetic and retrieved equilibrium fractionation in experiments, plotted against deposition-weighted supersaturation. Top halves of panels show kinetic effects from the diffusive model of [24] with a given value of d (circles), and with the default d=1.0164 of Cappa et al. (open diamonds) for reference. Bottom halves shows resulting equilibrium fractionations as computed slopes $\hat{a}_1^n$. (See Fig. S9.) Deviation from slope 0 implies a mis-specified kinetic model. Dashed lines show $a_1$ values corresponding to M67 (red) and this work (black). Blue line is weighted fit to $\hat{a}_1^n$s, excluding three outlier experiments (#4, 26, and 48, shown as open circles). The three panels show the fitted optimal value for d and conservative upper and lower bounds. Bounds span the range of published estimates of d.

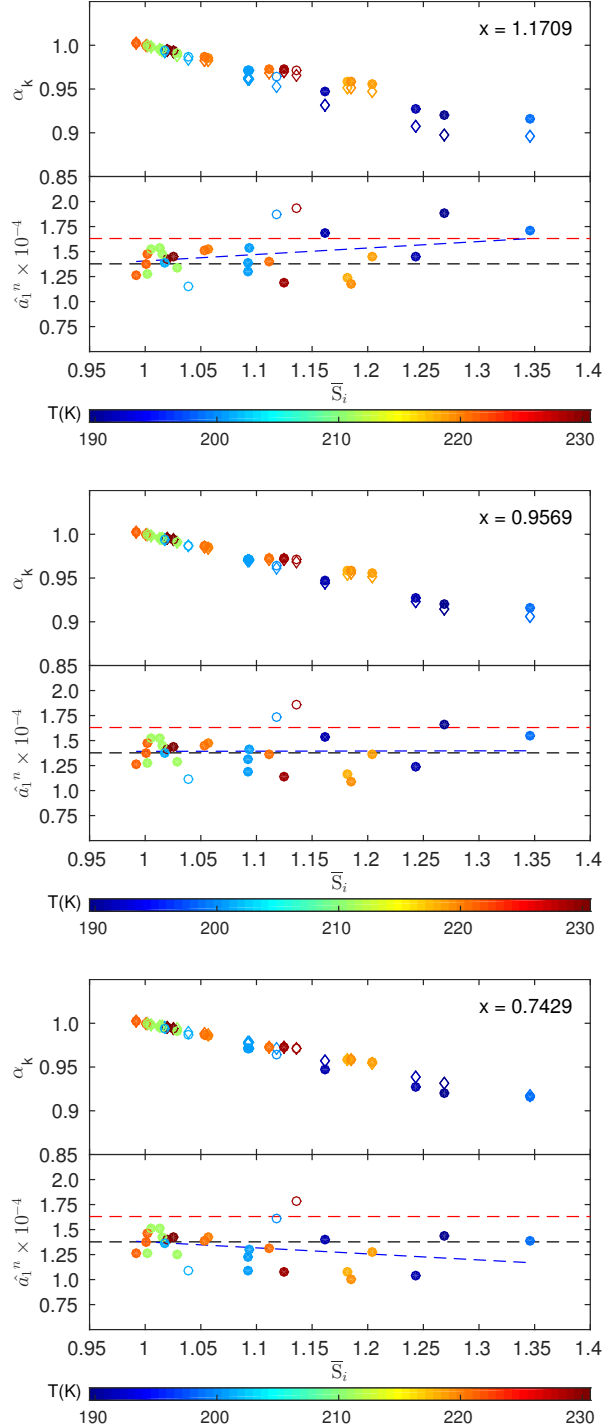

Figure S11: Effect of choice of x (ratio of isotopic deposition coefficients) on calculated kinetic and retrieved equilibrium fractionation in experiments, plotted against deposition-weighted supersaturation. Top halves of panels show kinetic effects from the diffusive/surface-kinetic model of [25] with a given value of x (circles), and with the diffusive model of [24] (open diamonds) for reference. In all cases d is set at 1.0164, the default value from Cappa et al. Bottom halves shows resulting equilibrium fractionations as computed slopes $\hat{a}_1^n$. (See Fig. S9.) Deviation from slope 0 implies a mis-specified kinetic model. Dashed lines show $a_1$ values corresponding to M67 (red) and this work (black). Blue line is weighted fit to $\hat{a}_1^n$s, excluding three outlier experiments (#4, 26, and 48, shown as open circles). The three panels show the fitted optimal value for x and conservative upper and lower bounds. Bounds span suggested range of plausible values for x.

## S7 Sensitivity tests on determination of $\alpha_{\text{eq}}$

We perform several tests to evaluate the robustness of our determination of $\alpha_{\text{eq}}$. We test for sensitivity to model assumptions (values for d), to measurement uncertainties (experimental region selection and bias in measured H$_2$O), and to fitting procedure (weighting of experiments in global fit). All factors produce relatively minor changes, and retain the finding of $\alpha_{\text{eq}}$ below M67.

### S7.1 Sensitivity to region choice

In all analyses described previously, we choose the experimental region analyzed for each pumpdown by a set of fixed criteria, described in S4:

- that ice deposition rate exceeds wall outgassing: $\epsilon = S_{wv}/P_{vi} < 1$
- that cumulative ice deposition exceeds a fraction $\Gamma$ of initial vapour, with $\Gamma = .02$
- that the isotopic ratio has dropped by a fraction $\Delta$ from its initial value, with $\Delta = 0.005$

The choice of region length can affect derived values of equilibrium fractionation for individual experiments ($\hat{\alpha}_n$) and therefore global fits for $\alpha_{\text{eq}}$, for several reasons. As discussed previously, wall outgassing isotopic composition is assumed constant but may show some trend over time. More importantly, chamber inhomogeneities produce fluctuations in measured quantities with timescales $\sim$30 s. With our region selection criteria, the timeseries analyzed for each pumpdown may contain 2-8 cycles of statistically significant fluctuations in measured quantities that affect $\hat{\alpha}_n$. Default values for $\epsilon$, $\Gamma$, and $\Delta$ were chosen such that derived values for $\hat{\alpha}_n$ are relatively stable to moderate extensions or reductions of region length. We evaluate here sensitivity to these parameter choices by performing a Bayesian analysis. We allow $\epsilon$, $\Gamma$, and $\Delta$ to vary arbitrarily (within bounds $\epsilon \in [0.3, 2.0]$, $\Gamma \in [0.01, 0.05]$ and $\Delta \in [0.001, 0.01]$), producing a wide range of candidate time segments. (In this analysis the region choice is randomly chosen and all experiments are evaluated with the same choice of parameters to define the region length.) We use each choice to derive an estimate for $\alpha_{\text{eq}}$, and examine all results that meet goodness-of-fit criteria.[1] Details of the analysis are described at the end of this section.

While the region choice parameters can strongly affect a resulting global fit for $\alpha_{\text{eq}}$, we find that the central value of the distribution of estimated $\hat{a}_1$ in this analysis is very close to that derived with the region choice algorithm of S4. That is, the most probable temperature-dependent $\alpha_{\text{eq}}$ in the Bayesian analysis is consistent with the determination of $\alpha_{\text{eq}}$ in our previous analyses (Fig. S12). (The confidence interval derived from the Bayesian analysis

---

[1] **Goodness of fit:** We evaluate the goodness of the fit by: first, checking that the minimizer has converged to a solution ($\nabla_{\boldsymbol{\theta}}\mathcal{L} \approx \mathbf{0}$ to specified tolerance and $\nabla_{\boldsymbol{\theta\theta}}\mathcal{L}$ positive definite); second, inspecting the normalized residuals

$$\frac{R_i^{\text{obs}} - R_i^{\text{m}}(\boldsymbol{\theta})}{\sigma_i}$$

and checking that their distribution does not significantly deviate from a Gaussian with mean 0 and variance 1 (to ensure compliance with Eq. (11)). This is assessed by a Kolmogorov-Smirnov test with a 5% significance level. Only fits to the randomly chosen regions which pass the goodness of fit test are included in the probability distribution.

is also smaller than that determined by the uncertainty analysis described in S5.) These results suggest that our region choice criteria have not biased our estimation of $\alpha_{eq}$.

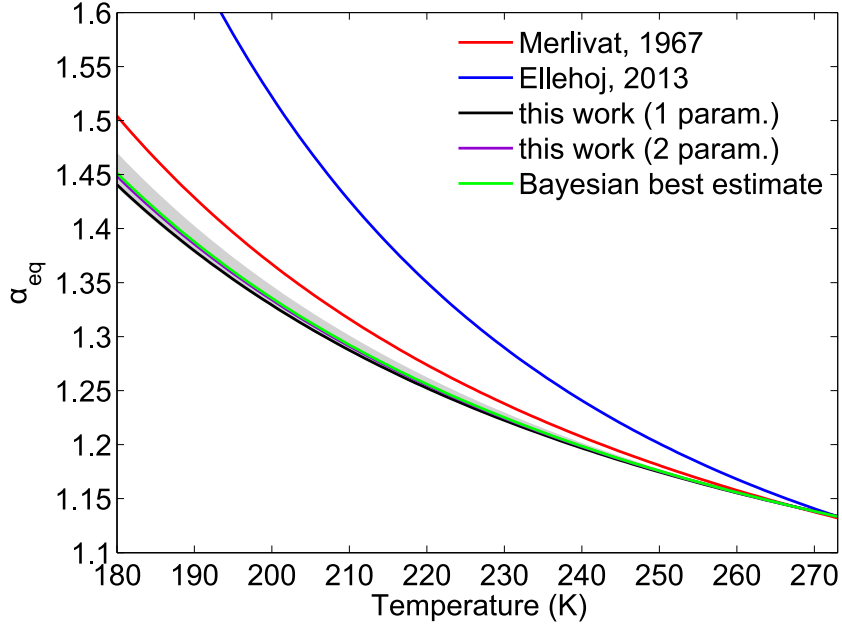

Figure S12: Test of sensitivity of global fit to region choice for individual experiments. We randomly vary parameters that determine the experiment region, then calculate the best global fit as described in the text (using a 2-parameter fit). The expectation value of the resulting probability distribution is shown in green. Gray shaded region is the 99.73% confidence interval derived from the probability distribution. Results suggest that our region choice has not introduced systematic bias. The Bayesian best estimate gives $a_0 = -0.0649$ and $a_1 = 14171$, which is very close to the best 2-parameter fit with the region choice used in the analysis. (In this test we take d from Merlivat, 1978, but d negligibly affects results; see S7.2 below.)

*Bayesian analysis methods*: The estimator of $a_1$ computed in Eq. (18) implicitly depends on the chosen values of $\epsilon$, $\Gamma$, and $\Delta$ and can be seen as deterministic when these parameters are fixed. In this analysis we treat $\hat{a}_1$ as a random variable with a PDF related to probability distributions of $\epsilon$, $\Gamma$, and $\Delta$:

$$p(\hat{a}_1) = \iiint_{-\infty}^{+\infty} p(\hat{a}_1|\epsilon, \Gamma, \Delta) \, p(\epsilon, \Gamma, \Delta) \, d\epsilon \, d\Gamma \, d\Delta$$
$$= \iiint_{-\infty}^{+\infty} \boldsymbol{\delta}(\hat{a}_1 - \hat{a}_1(\epsilon, \Gamma, \Delta)) \, p(\epsilon, \Gamma, \Delta) \, d\epsilon \, d\Gamma \, d\Delta \qquad (26)$$

where $\boldsymbol{\delta}$ represents the Dirac distribution. If $\epsilon$, $\Gamma$, and $\Delta$ are independent and uniformly distributed between bounds, we can draw $n$ samples (with $n$ sufficiently large; we use $n = 100$) and evaluate the density of $\hat{a}_1$ as:

$$p(\hat{a}_1) \approx \frac{1}{n} \sum_{k=1}^{n} \boldsymbol{\delta}(\hat{a}_1 - \hat{a}_1(\epsilon_k, \Gamma_k, \Delta_k)) \qquad (27)$$

or, more conveniently, to give a continuous representation of the density:

$$p(\hat{a}_1) \approx \frac{1}{n} \sum_{k=1}^{n} \mathbf{G}\left(\hat{a}_1 - \hat{a}_1\left(\epsilon_k, \Gamma_k, \Delta_k\right)\right) \tag{28}$$

where $\mathbf{G}$ is a Gaussian kernel of suitable width. The most probable $\hat{a}_1$ and confidence intervals on the estimate can then be computed from the probability distribution $p(\hat{a}_1)$.

### S7.2 Sensitivity to estimation of kinetic isotope effects

In this section we test the robustness of our estimation of $\alpha_{\text{eq}}$ to uncertainties that affect the calculated kinetic modification to fractionation: uncertainty in the isotopic diffusivity ratio d and potential systematic bias in the measurement of $H_2O$. Resulting errors in calculated kinetic effects could affect the global estimate of $\alpha_{\text{eq}}$.

*Isotopic diffusivity ratio (d)*
In section S6, we showed that different assumed values of d could significantly alter the slope of inferred $\hat{\alpha}_n$ against supersaturation. Those changes have a much smaller effect on the global fit for $\alpha_{\text{eq}}$, because the global fit of the IsoCloud 4 dataset is dominated by experiments with small supersaturations and minimal kinetic effects. Varying d over the range of published estimates (∼3% variation, see Table S5) can affect estimation of $\hat{\alpha}_n$ in individual experiments by as much as 1-2% (Figure S13), but has virtually no effect on the retrieval of $\alpha_{\text{eq}}$, producing a maximum change of 0.02%. (We do not show a figure for this test because the differences in $\alpha_{\text{eq}}$ are too small to be visible by eye.) Uncertainty in d therefore cannot bias our estimate of $\alpha_{\text{eq}}$.

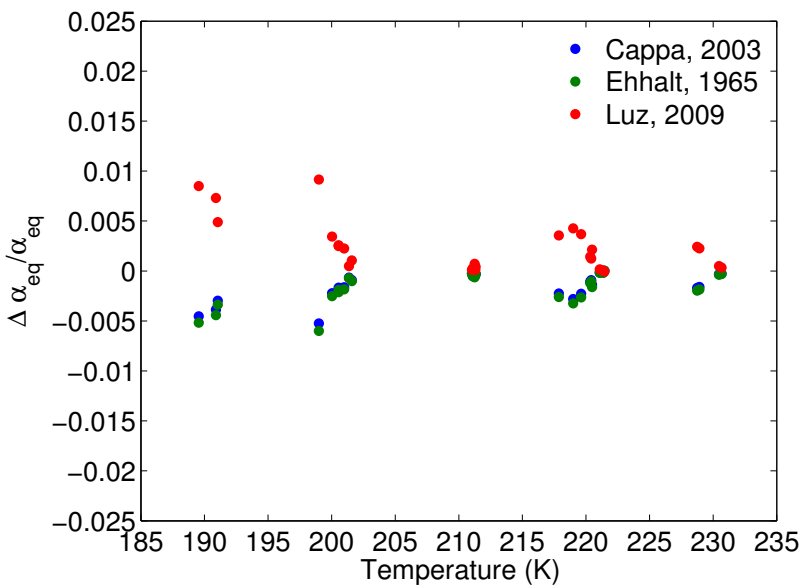

Figure S13: Effect of choices for the isotopic diffusivity ratio d on inferred $\hat{\alpha}_n$ for individual experiments. We use the value of d from Merlivat 1978 as the reference and show differences when d is taken instead from Cappa et al. 2003 (blue), Ehhalt and Knott 1965 (green), and Luz et al. 2009 (red). Individual experiments may be affected by up to 1-2%, but the global fit for the temperature dependence of $\alpha_{\text{eq}}$ must intersect the cluster of near-equilibrium experiments at ∼210 K that show negligible dependence on d.

*Supersaturation*

Systematic bias in $H_2O$ measurements can affect estimates of isotopic fractionation in several ways: by changing the inferred ice deposition rate, and by producing a biased measurement of supersaturation $S_i$ and so a biased calculation of the kinetic modification to fractionation. The latter effect is by far the most important. Systematic measurement bias is common for measurements of gas-phase species by absorption spectroscopy that do not involve empirical calibration, since spectral line parameters are not perfectly known. For the ChiWIS $H_2O$ measurements, stated uncertainties on line parameters from the HITRAN database [14, 15] nominally limit the water vapour retrieval to an accuracy of $\pm 5\%$. However, ChiWIS measured vapour pressure in dense ice clouds, when water vapour should be drawn down to saturation, suggests that ChiWIS $H_2O$ measurements are accurate to within $-1\%$ and $+2\%$. (These limits are 2 standard deviations around the mean of calculated ChiWIS $S_i$, using [37] for saturation vapour pressure.) Potential biases of this magnitude would produce noticeable changes in estimated $\alpha_{eq}$, since experiments at $S_i \sim 1$ would then be assumed to experience super- or sub-saturation and kinetic modifications to fractionation. The effect is shown in Fig. S14. Systematic over- or under-estimation of supersaturation is the largest source of systematic uncertainty in our analysis. However, the range of potential changes remains within the $3\sigma$ confidence interval of the best fit presented here.

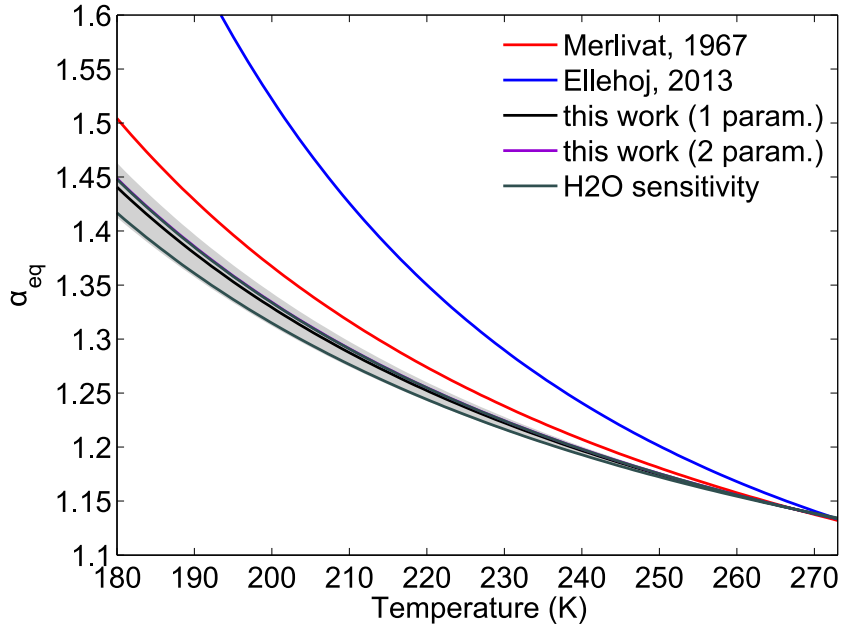

Figure S14: Sensitivity of fit for $\alpha_{eq}$ to potential fractional bias in water vapour and therefore $S_i$. Black line shows default best-fit $\alpha_{eq}$; grey lines repeat analysis with assumption that ChiWIS $H_2O$ has bias of $+2\%$ and $-1\%$. Overestimating $H_2O$ would lead to overestimating kinetic modification to fractionation and so $\alpha_{eq}$; the true value of $\alpha_{eq}$ would then be lower than estimated. Similary, underestimating $H_2O$ would mean the true value of $\alpha_{eq}$ would be higher than estimated. Gray shaded region is the $3\sigma$ confidence interval described previously.

## S7.3 Sensitivity to weights on individual experiments

As mentioned previously, our estimated experimental uncertainties may be subject to temperature-dependent bias. The global fit procedure, which weights individual experiments by their estimated error, may therefore over-weight the warm-temperature experiments relative to those at $T < 210$ K. We test for any resulting bias in determination of $\alpha_{eq}$ by comparing to a fit with all experiments treated equally. The parametrizations for $\alpha_{eq}$ in the weighted and unweighted fits are not significantly different (Figure S15). (The parameters in the unweighted fit are $a_0 = -0.0595$ and $a_1 = 13837$; compare to values in Table S4.) Our estimate of $\alpha_{eq}$ appears robust to errors in the weighting of individual experiments.

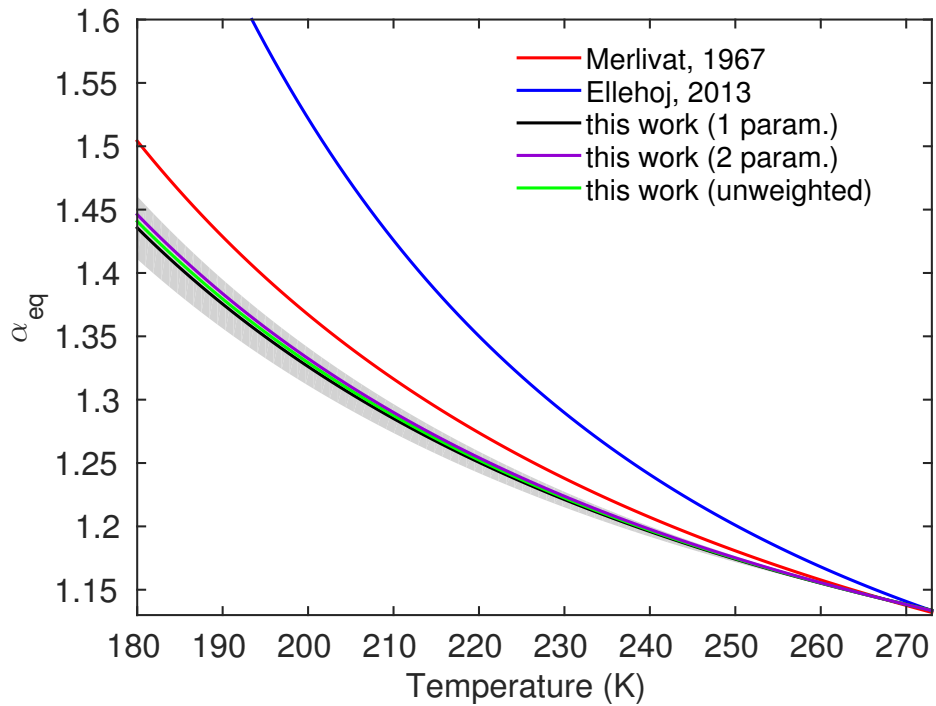

Figure S15: Sensitivity of the fit for $\alpha_{eq}$ to potential uncertainty in the weighting of individual experiments. Black line shows default best-fit $\alpha_{eq}$ using weights derived from the uncertainty analysis; green line repeats the analysis with equal weighting for all experiments. Gray shaded region is the $3\sigma$ confidence interval described previously.


\end{document}